\documentclass[11pt]{article}

\usepackage{epsfig,subfigure}
\usepackage{amssymb, amsmath}
\usepackage{color}
\usepackage{graphicx}

\usepackage{amssymb}
\usepackage{amsmath,amsfonts,amssymb}
\usepackage{verbatim}
\usepackage{stfloats}
\usepackage[bookmarks=false]{}

\newtheorem{theorem}{Theorem}

\newtheorem{lemma}{Lemma}

\newtheorem{example}{Example}

\newcommand\blfootnote[1]{%
  \begingroup
  \renewcommand\thefootnote{}\footnote{#1}%
  \addtocounter{footnote}{-1}%
  \endgroup
}

\begin{document}
\date{}
\title{On the Optimality of Treating Interference as Noise:\\
General Message Sets}
\author{ \normalsize Chunhua Geng, Hua Sun and Syed A. Jafar  \\
}

\maketitle
\blfootnote{Chunhua Geng (email: chunhug@uci.edu), Hua Sun (email: huas2@uci.edu) and Syed A. Jafar (email: syed@uci.edu) are with the Center of Pervasive Communications and Computing (CPCC) in the Department of Electrical Engineering and Computer Science (EECS) at the University of California Irvine.}

\begin{abstract}
In a $K$-user Gaussian interference channel, it has been shown that if for each user the desired signal strength is no less than the sum of the strengths of the strongest interference from this user and the strongest interference to this user (all values in dB scale), then treating interference as noise (TIN) is optimal from the perspective of generalized degrees-of-freedom (GDoF) and achieves the entire channel capacity region to within a constant gap. In this work, we  show that for such TIN-optimal interference channels, even if the message set is expanded to include an independent message from each transmitter to each receiver,  operating the new channel as the original interference channel and treating interference as noise is still optimal for the \emph{sum}  capacity up to a constant gap. Furthermore, we extend the  result to the sum-GDoF optimality of TIN in the general setting of $X$ channels with arbitrary numbers of transmitters and receivers.

\end{abstract}


\section{Introduction}
Treating interference as noise (TIN) when it is sufficiently weak is an attractive interference management principle for
wireless networks in practice due to its simplicity and robustness. Remarkably, TIN is also information-theoretically optimal
when the interference is sufficiently weak. This is established  in
\cite{Motahari_Khandani_TIN, Annapureddy_Veeravalli_TIN_opt, Kramer_TIN_opt, Annapureddy_Veeravalli_MIMO, Shang_Kramer_parallel, Shang_Kramer_vector}
from an exact capacity perspective, and in  \cite{Tse_GDoF,Jafar_Vishwanath_GDoF, Gou_Jafar_O1, Karmakar_Varanasi_GDoF, Karmakar_Varanasi_gap, Geng_TIN_opt}
from an approximate capacity perspective. Each approach has its merits -- the former identifies relatively narrow regimes where TIN
achieves exact capacity, whereas the latter identifies significantly broader regimes where TIN is approximately optimal. Most relevant
to this work are the results by Geng et al. in \cite{Geng_TIN_opt} where it is shown that in a general $K$-user interference channel,
if  for each user the desired signal strength is no less than the sum of the strengths of the strongest interference \emph{from} this
user and the strongest interference \emph{to} this user (all values in dB scale), then TIN is optimal for the entire channel capacity region
up to a constant gap of no more than $\log_2(3K)$ bits. 

In this paper we explore the sum-rate optimality of TIN when the message set is expanded to include an independent message from each transmitter to each receiver, i.e., the $X$ channel setting \cite{Jafar_Shamai_X,Maddah_X,Cadambe_Jafar_X}. Related prior works on the $X$ setting in \cite{Huang_GDoF_X, Niesen_Maddah} have primarily focused on the case with 2 transmitters and 2 receivers. In \cite{Huang_GDoF_X}, Huang, Cadambe and Jafar characterize the sum-GDoF for the symmetric $X$ channel and identify sufficient conditions for TIN to achieve exact capacity in the asymmetric case. In \cite{Niesen_Maddah}, Niesen and Maddah-Ali characterize the capacity for the general asymmetric case within a constant gap subject to an outage set.

The main contribution of this work is to show that, for the $K$-user TIN-optimal interference channels identified by Geng et al. in \cite{Geng_TIN_opt}, even if the message set is expanded to also include an independent message from each transmitter to each receiver, operating as the original interference channel and treating interference as noise at each receiver is still optimal for the \emph{sum}  capacity up to a constant gap (see Theorem \ref{tin-X-GDoF} in Section \ref{sec_res}). We also extend the optimality of TIN to the general $X$ channel with arbitrary numbers of transmitters and receivers (see Theorem \ref{tin-X-asy} in Section \ref{sec_res}). Notably, to complete the generalization to $X$ channels, we resort to  deterministic channel models, and use the fact that the sum capacity of Gaussian channels is upper bounded by that of their carefully chosen deterministic counterparts up to a constant gap. 

\section{Preliminaries}

\subsection{Channel Model}

Consider the wireless channel with $M$ transmitters and $N$ receivers, which can be described by the following input-output equations,
\begin{equation}
\label{ori_channel}
Y_k(t)=\sum_{i=1}^M\tilde{h}_{ki}\tilde{X}_i(t)+Z_k(t),~~~\forall k\in\{1,2,...,N\},
\end{equation}
where $\tilde{h}_{ki}$ is the complex channel gain value from transmitter $i$ to receiver $k$. $\tilde{X}_i(t)$, $Y_k(t)$ and $Z_k(t)$ are the transmitted symbol of transmitter $i$, the received signal of receiver $k$, and the additive circularly symmetric complex Gaussian noise with zero mean and unit variance seen by receiver $k$, respectively, at each time index $t$. All the symbols are complex. Each transmitter $i$ is subject to the power constraint $E[|\tilde{X}_i(t)|^2]\leq P_i$.

Following  similar approaches in \cite{Tse_GDoF,Geng_TIN_opt}, we translate the standard channel model (\ref{ori_channel}) into an equivalent normalized form that is more conducive for GDoF studies. We define
\begin{align}
\label{ratio}
\alpha_{ki}\triangleq\frac{\log(\max\{1,|\tilde{h}_{ki}|^2P_i\})}{\log P},~~\forall i\in\{1,2,...,M\}, \forall k\in\{1,2,...,N\},
\end{align}
where $P>1$ is a nominal power value.

Now according to (\ref{ratio}), we represent the original channel model (\ref{ori_channel}) in the following form,
\begin{equation}
\label{equ_channel}
\begin{aligned}
Y_k(t)=&\sum_{i=1}^Mh_{ki}X_i(t)+Z_k(t)\\
=&\sum_{i=1}^M\sqrt{P^{\alpha_{ki}}}e^{j\theta_{ki}}X_i(t)+Z_k(t),~~~ \forall k\in\{1,2,...,N\}.
\end{aligned}
\end{equation}
where $X_i(t)=\tilde{X}_i(t)/\sqrt{P_i}$ is the normalized transmit symbol of transmitter $i$, subject to the unit power constraint, i.e., $E[|X_i(t)|^2]\leq 1$. $\sqrt{P^{\alpha_{ki}}}$ and $\theta_{ki}$ are the magnitude and the phase, respectively, of the channel between transmitter $i$ and receiver $k$. The exponent $\alpha_{ki}$ is called the channel strength level of the link between transmitter $i$ and receiver $k$. As in \cite{Tse_GDoF,Geng_TIN_opt}, for the GDoF metric, we preserve the ratios $\alpha_{ki}$ as all SNRs approach infinity. In the rest of the paper, we only consider the equivalent channel model in (\ref{equ_channel}).

In the $K$-user interference channel where $M=N=K$, each transmitter intends to send one independent message to its corresponding receiver. Because we wish to prove the negative result that additional messages do not add to the sum-GDoF in a TIN-optimal network, the strongest result corresponds to the case where we include messages from every transmitter to every receiver. Therefore, we will consider the $X$ channel setting. In the $M\times N$ $X$ channel, transmitter $i$ has message $W_{ki}$ intended for receiver $k$, and  the messages $\{W_{ki}\}$ are independent, $\forall i\in\{1,2,...,M\}, \forall k\in\{1,2,...,N\}$.  The size of the message set $\{W_{ki}\}$ is denoted by $|W_{ki}|$. For codewords spanning $n$ channel uses, the rates $R_{ki}=\frac{\log|W_{ki}|}{n}$ are achievable if the probability of error of all messages can be made arbitrarily small simultaneously by choosing an appropriately large $n$. The channel capacity region $\mathcal{C}$ is the closure of the set of all achievable rate tuples. Collecting the channel strength levels and phases in the sets
\begin{equation}
\alpha\triangleq\{\alpha_{ki}\},~~~\theta\triangleq\{\theta_{ki}\},~~~\forall i\in\{1,2,...,M\},~ \forall k\in\{1,2,...,N\},
\end{equation}
the capacity region  is denoted as $\mathcal{C}(P,\alpha,\theta)$, which is a function of $\alpha$, $\theta$, and $P$. The sum channel capacity is defined as
\begin{align}
C_{\Sigma,X}=\max_{\mathcal{C}(P,\alpha,\theta)}\sum_{i=1}^M\sum_{k=1}^NR_{ki}
\end{align}
Then the GDoF region of the $X$ channel as represented in (\ref{equ_channel}) is given by
\begin{equation}
\begin{aligned}
\mathcal{D}(\alpha,\theta)\triangleq \Big\{(d_{11},d_{12},...,d_{NM}): ~&d_{ki}=\lim_{P\rightarrow\infty}\frac{R_{ki}}{\log P},~~~\forall i\in\{1,2,...,M\}, \forall k\in\{1,2,...,N\},\\
&(R_{11},R_{12},...,R_{NM})\in \mathcal{C}(P,\alpha,\theta)\Big\},
\end{aligned}
\end{equation}
and its sum-GDoF value is
\begin{align}
d_{\Sigma,X}=\max_{\mathcal{D}(\alpha,\theta)}\sum_{i=1}^M\sum_{k=1}^Nd_{ki}
\end{align}


\subsection{On the Optimality of TIN for Interference Channel}
Let us first review the optimality of TIN for the $K$-user interference channel from the perspective of GDoF.


\begin{theorem}\label{tin-IC}
\emph{(Theorem 1 in \cite{Geng_TIN_opt})}
In a $K$-user interference channel, where the channel strength level from transmitter $i$ to receiver $j$ is equal to $\alpha_{ji}$, $\forall i,j\in\{1,...,K\}$, if the following condition is satisfied
\begin{align}\label{cond_ic}
\alpha_{ii}\geq \max_{j:j\neq i}\{\alpha_{ji}\}+\max_{k:k\neq i}\{\alpha_{ik}\},~~~\forall i,j,k\in\{1,2,...,K\},
\end{align}
then power control and treating interference as noise  achieves the whole GDoF region. Moreover, the GDoF region is the set of all $K$-tuples $(d_1,d_2,...,d_K)$ satisfying
\begin{alignat}{2}
\mbox{individual bounds: } 0\leq d_i&\leq \alpha_{ii},\:\:&&\forall i\in\{1,...,K\}\label{eqq11}\\
\mbox{cycle bounds: } \sum_{j=1}^{m}d_{i_j}&\leq \sum_{j=1}^{m}(\alpha_{i_ji_j}-\alpha_{i_{j-1}i_j}),\:\:&&\forall (i_1,...,i_m)\in\Pi_K,~~\forall m\in\{2,3,...,K\},\label{eqq12}
\end{alignat}
where $\Pi_K$ is the set of all possible cyclic sequences of all subsets of $\{1,...,K\}$, and the modulo-$m$ arithmetic is implicitly used on the user indices, e.g., $i_m=i_0$.
\end{theorem}

\emph{Remark}: The above theorem claims that in the $K$-user interference channel, if \emph{for each user the desired signal strength is no less than the sum of  the strengths of the strongest interference from this user and the strongest interference to this user (all values in dB scale)}, then TIN is GDoF-optimal. Furthermore, it is shown in  \cite{Geng_TIN_opt} that under the same condition, TIN achieves the entire channel capacity region to within a  gap no larger than $\log_2(3K)$ bits. Note that the gap is bounded by a constant for a fixed number of users, i.e., it does not depend on the channel strength parameters $\alpha_{ij}$ and $P$.



\section{Results} \label{sec_res}
The main result of this paper is the following theorem.

\begin{theorem}
\label{tin-X-GDoF}
In a $K$-user interference channel, where the channel strength level from transmitter $i$ to receiver $j$ is equal to $\alpha_{ji}$, $\forall i,j\in\{1,2,...,K\} $, when the following condition is satisfied,
\begin{align}\label{cond}
\alpha_{ii}\geq \max_{j:j\neq i}\{\alpha_{ji}\}+\max_{k:k\neq i}\{\alpha_{ik}\}~~\forall i,j,k\in\{1,2,...,K\}
\end{align}
then even if the message set is  increased to the $X$ channel setting, operating the new channel as the original interference channel  and treating interference as noise at each receiver still achieves the sum-GDoF. Furthermore, the same scheme is also optimal for the sum channel capacity up to a constant gap of no more than $K\log_2[K(K+1)]$ bits.
\end{theorem}


%

The proof of Theorem \ref{tin-X-GDoF} is presented in Section \ref{proofs}.

While the $K$-user interference channel is naturally associated with a $K\times K$ $X$ channel setting, the $X$ channel setting also allows for unequal numbers of transmitters and receivers. A natural question is whether such a generalization of the TIN-optimality result is possible for  the $X$ channel with $M\neq N$. The following theorem provides such a generalization.



\begin{theorem}
\label{tin-X-asy}
In an $M\times N$ $X$ channel, where the channel strength level from transmitter $i$ to receiver $j$ is equal to $\alpha_{ji}$ and $\kappa\triangleq \min\{M,N\}$, if there exist two permutations $\Pi^T$ and $\Pi^R$ for the transmitter and receiver indices, respectively, such that
\begin{align}\label{cond_asy}
\alpha_{\Pi^R_i\Pi^T_i}\geq \max_{j:j\neq i}\{\alpha_{\Pi^R_j\Pi^T_i}\}+\max_{k:k\neq i}\{\alpha_{\Pi^R_i\Pi^T_k}\}~~\forall i\in\{1,2,...,\kappa\},\forall j\in\{1,2,...,N\},\forall k\in\{1,2,...,M\},
\end{align}
where $\Pi^T_i$ ($\Pi^R_i$) denotes the $i$-th element in the permutation of transmitters (receivers) $\Pi^T$ ($\Pi^R$), then operating the channel as a $\kappa$-user interference channel and treating interference as noise at each receiver is sum-GDoF optimal\footnote{Based on the proof in Section \ref{proofs2}, it is not hard to verify that the same TIN scheme is also optimal to achieve the sum channel capacity to within a constant gap.}.
\end{theorem}



\begin{figure}[h]
\begin{center}
 \includegraphics[width= 7 cm]{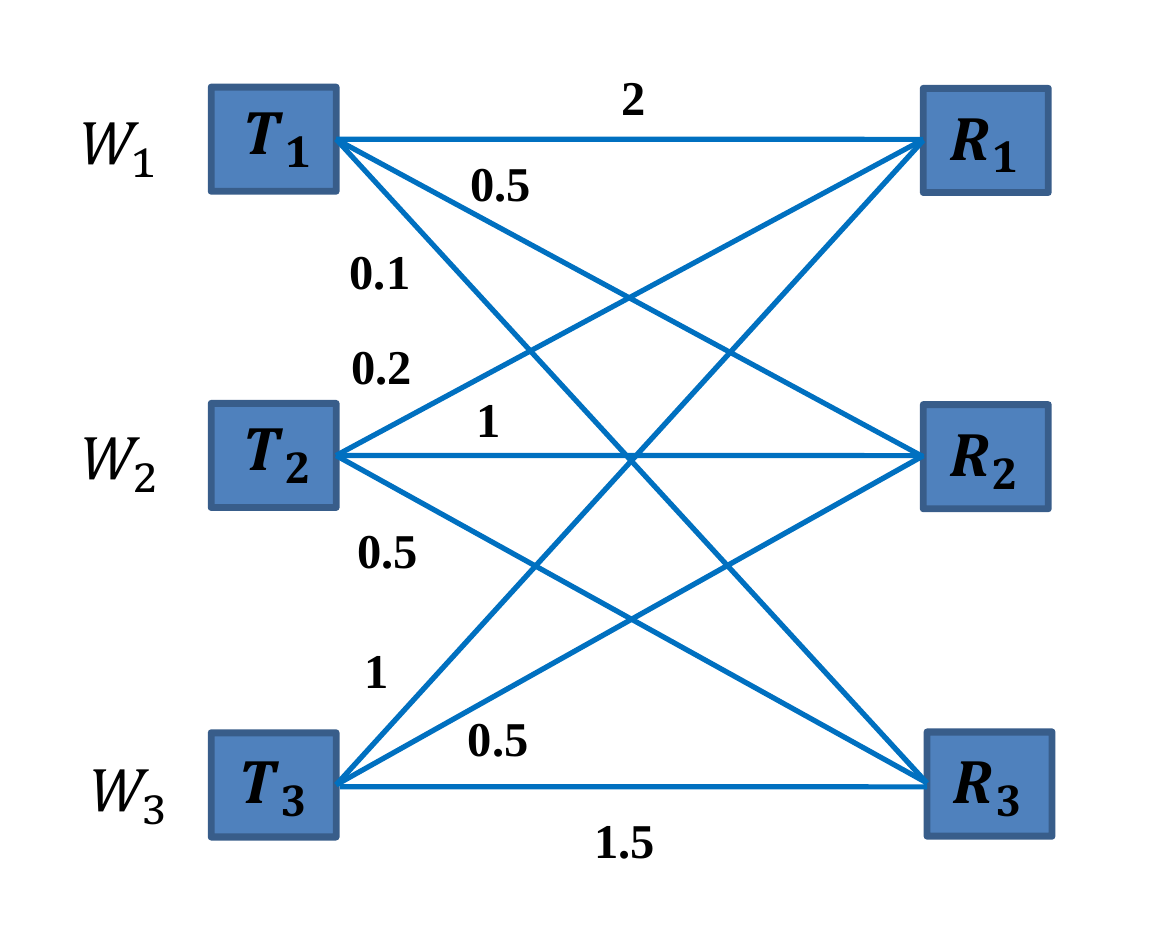}
 \caption{A 3-user interference channel, where the value on each link denotes its channel strength level.}
\label{3user_IC}
\end{center}
\end{figure}

\begin{example}
First, consider the $3$-user interference channel illustrated in Fig.~\ref{3user_IC}, where transmitter $i$ intends to send an independent message to its desired receiver $i$, $\forall i\in\{1,2,3\}$. Note there are $3$ messages in this setting. It's easy to check that the TIN-optimal condition (\ref{cond}) is satisfied for each user. Then according to Theorem \ref{tin-IC}, it is not hard to verify that the sum-GDoF value of this interference channel is
\begin{align*}
d_{\Sigma,IC}=d_1+d_2+d_3=2.5
\end{align*}
which is achieved by power control and TIN.

\begin{figure}[h]
\begin{center}
 \includegraphics[width= 9 cm]{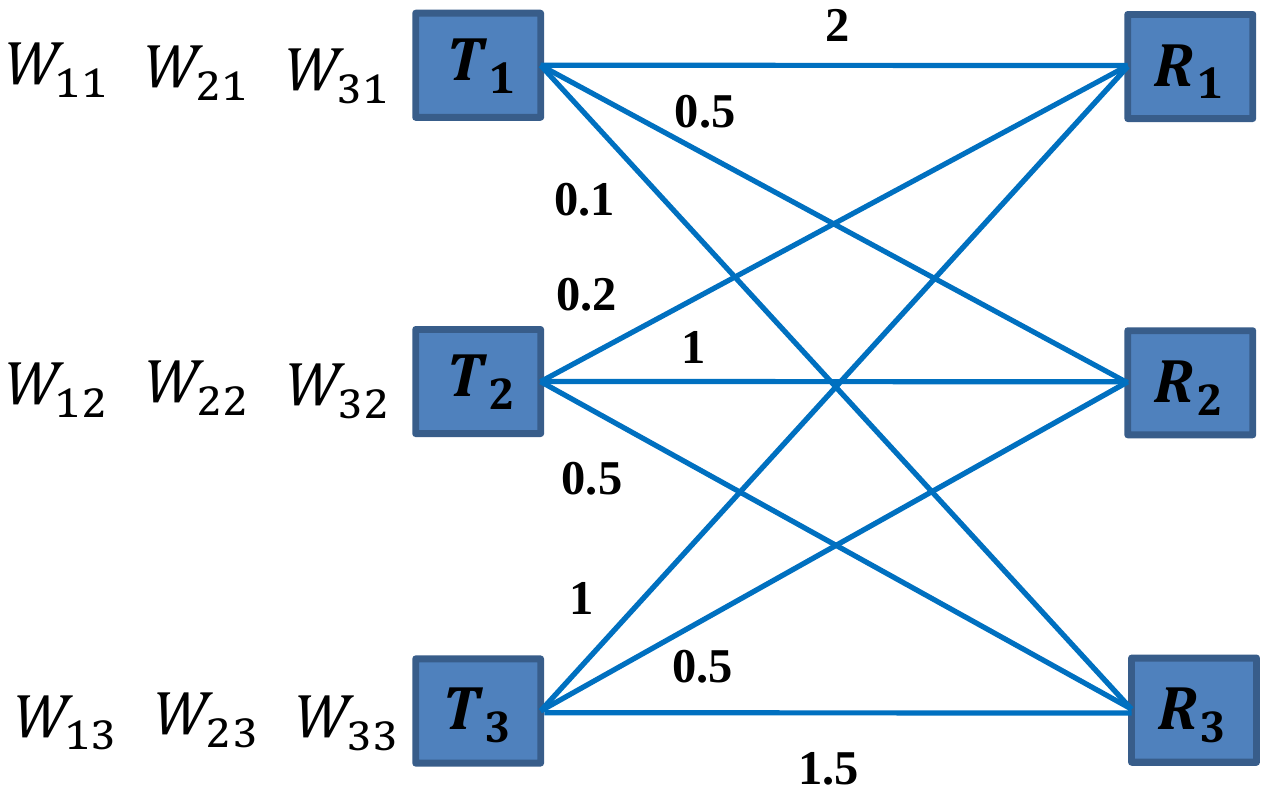}
 \caption{A $3\times 3$ $X$ channel, which has the same channel strength levels as the $3$-user interference channel in Fig.~\ref{3user_IC}.}
\label{3user_X}
\end{center}
\end{figure}

Next, let us expand the set of messages to the $X$ channel setting, where each transmitter intends to send an independent message to each receiver as shown in Fig.~\ref{3user_X}. Therefore, there are totally $9$ messages in this $X$ channel. Theorem \ref{tin-X-GDoF} claims that for this $3\times 3$ X channel, the sum-GDoF value is still
\begin{align*}
d_{\Sigma,X}=\sum_{i=1}^3\sum_{k=1}^3d_{ki}=2.5
\end{align*}
which can be achieved by setting $W_{ki}=\phi$ for $i\neq k$ and $\forall i,k\in\{1,2,3\}$, sending only $\{W_{11},W_{22},W_{33}\}$ through the channel and treating interference as noise at each receiver.

\begin{figure}[h]
\begin{center}
 \includegraphics[width= 10 cm]{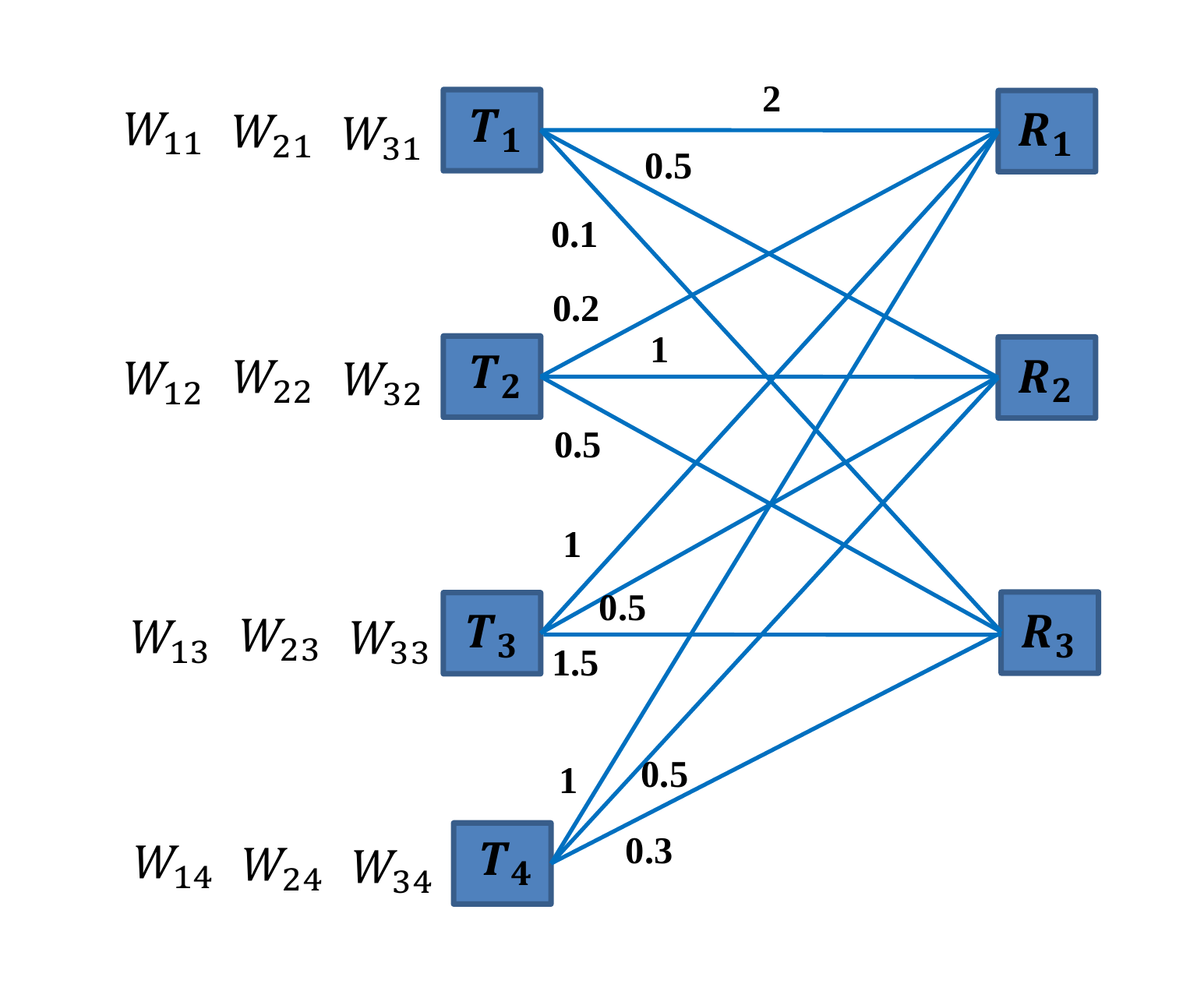}
 \caption{A $4\times 3$ $X$ channel, which is obtained by adding another transmitter to the $X$ channel in Fig.~\ref{3user_X}.}
\label{4-3user_X}
\end{center}
\end{figure}

Finally, after adding another transmitter as depicted in Fig.~\ref{4-3user_X}, the number of the messages increases to $12$ in this $4\times 3$ $X$ channel. It's easy to verify that (\ref{cond_asy}) holds. Then according to Theorem \ref{tin-X-asy}, for this $X$ channel and its reciprocal channel, the same TIN scheme  is still optimal in terms of the sum-GDoF, whose value remains $2.5$.

\end{example}

\section{Proofs}\label{proofs}

\subsection{Proof for Theorem \ref{tin-X-GDoF}}
In the following, we first consider the sum-GDoF of the $K\times K$ $X$ channel. Then we use the insight gained in the GDoF study to derive the constant gap result for the sum channel capacity.

\bigskip
\textbf{Proof for the sum-GDoF}: The proof consists of two steps. In the first step, we show that for all individual and cycle bounds of a TIN-optimal $K$-user interference channel (see Theorem \ref{tin-IC}), if each $d_i$ ($\forall i\in\{1,2,...,K\}$) is replaced by $\hat{d}_i=\sum_{j=1}^Kd_{ij}$, these bounds still hold for its counterpart $X$ channel.

In the following, we first give an example of the $3\times 3$ $X$ channel, then generalize the proof to the $K\times K$ $X$ channel.

\begin{example}\label{ex2}
Consider a $3$-user TIN-optimal interference channel. According to Theorem \ref{tin-IC}, we can obtain the entire GDoF region, which is characterized by certain individual and cycle bounds. To extend the result to the X channel setting, each of these bounds will be extended. To illustrate the key ideas in this example, we consider the following two bounds,
\begin{align}
d_3&\leq \alpha_{33},\label{ex2_IC_ind}\\
d_1+d_2&\leq (\alpha_{11}+\alpha_{22})-(\alpha_{12}+\alpha_{21}),\label{ex2_IC_cycle}
\end{align}
and intend to prove that in the counterpart $3\times 3$ $X$ channel, if we replace each $d_i$ by $\hat{d}_i=\sum_{j=1}^3d_{ij}$, $\forall i\in\{1,2,3\}$, the above two bounds still hold, i.e.,
\begin{align}
\hat{d}_3=d_{31}+d_{32}+d_{33}&\leq \alpha_{33}\label{ex2_X_ind}\\
\hat{d}_1+\hat{d}_2=d_{11}+d_{12}+d_{13}+d_{21}+d_{22}+d_{23}&\leq (\alpha_{11}+\alpha_{22})-(\alpha_{12}+\alpha_{21})\label{ex2_X_cycle}
\end{align}
All the remaining bounds can be extended to the X channel similarly.

To prove (\ref{ex2_X_ind}), we just need to consider the MAC consisting of all the transmitters and the receiver $3$, then we have
\begin{align}
R_{31}+R_{32}+R_{33}\leq \log_2(1+P^{\alpha_{31}}+P^{\alpha_{32}}+P^{\alpha_{33}})
\end{align}
Because (\ref{cond}) is satisfied, i.e., $\alpha_{33}\geq \alpha_{32}$ and $\alpha_{33}\geq \alpha_{31}$, therefore in the GDoF sense we have
\begin{align}
\hat{d}_3=d_{31}+d_{32}+d_{33}\leq\alpha_{33}
\end{align}

\begin{figure}[h]
\begin{center}
 \includegraphics[width= 7 cm]{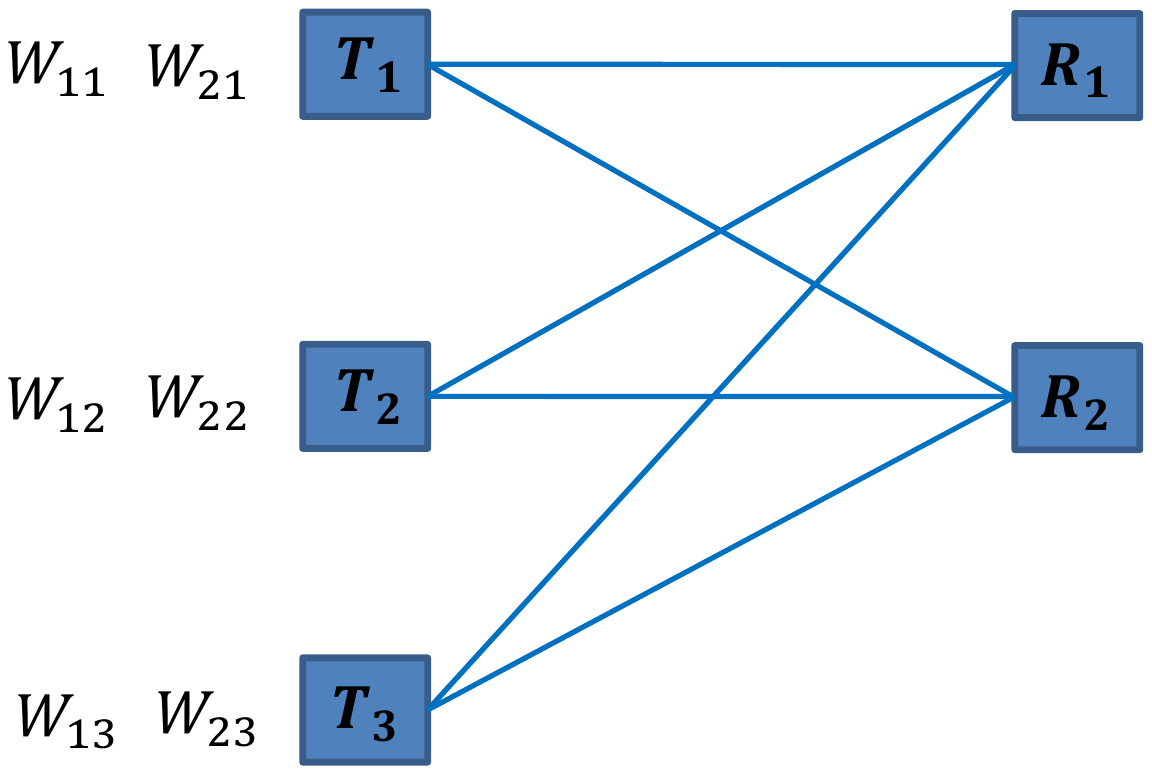}
 \caption{The subnetwork with 3 transmitters, 2 receivers and 6 messages}
\label{3-2user_X}
\end{center}
\end{figure}

To prove (\ref{ex2_X_cycle}), consider the subnetwork consisting of all the transmitters and the receivers $1$ and $2$ as illustrated in Fig.~\ref{3-2user_X}, where we have eliminated  the third receiver and its desired messages $W_{31}, W_{32}, W_{33}$. This cannot hurt the rates of the remaining messages, so the outer bound arguments remain valid. Define
\begin{align}
S_1(t)&=h_{21}X_1(t)+Z_2(t)\\
S_2(t)&=h_{12}X_2(t)+Z_1(t)
\end{align}
For  receiver $1$, we provide $S_1^n$, $W_{21}$ and $W_{23}$ through a genie. From Fano's inequality, we have
\begin{align}
&n(R_{11}+R_{12}+R_{13}-\epsilon)\\
&\leq I(W_{11},W_{12},W_{13};Y_1^n,S_1^n,W_{21},W_{23})\\
&=I(W_{11},W_{12},W_{13};Y_1^n,S_1^n|W_{21},W_{23}\label{ex2_S1})\\
&=I(W_{11},W_{12},W_{13};S_1^n|W_{21},W_{23})+I(W_{11},W_{12},W_{13};Y_1^n|S_1^n,W_{21},W_{23})\\
&=h(S_1^n|W_{21},W_{23})-h(S_1^n|W_{21},W_{23},W_{11},W_{12},W_{13})\nonumber\\
&~~~+h(Y_1^n|S_1^n,W_{21},W_{23})-h(Y_1^n|S_1^n,W_{21},W_{23},W_{11},W_{12},W_{13})\\
&\leq h(S_1^n|W_{21},W_{23})-h(S_1^n|W_{21},W_{23},W_{11},W_{12},W_{13},X_1^n)\nonumber\\
&~~~+h(Y_1^n|S_1^n,W_{21},W_{23})-h(Y_1^n|S_1^n,W_{21},W_{23},W_{11},W_{12},W_{13},X_1^n, X_3^n)\label{ex2_S2}\\
&\leq h(S_1^n|W_{21})-h(Z_2^n)+h(Y_1^n|S_1^n)-h(S_2^n|W_{12})\label{ex2_S3}
\end{align}
where (\ref{ex2_S1}) follows because all the messages are independent, (\ref{ex2_S2}) holds since adding conditioning does not increase entropy  and  (\ref{ex2_S3}) holds because dropping conditioning (in the first and third terms) does not reduce entropy.

Due to symmetry, for the receiver $2$, we similarly obtain
\begin{align}
n\left(R_{21}+R_{22}+R_{23}-\epsilon\right)&\leq h(S_2^n|W_{12})-h(Z_1^n)+h(Y_2^n|S_2^n)-h(S_1^n|W_{21})
\end{align}

Thus the sum rate is bounded as follows.
\begin{align}
n(\sum_{i=1}^2\sum_{j=1}^3R_{ij}-2\epsilon)&\leq h(Y_1^n|S_1^n)+h(Y_2^n|S_2^n)-h(Z_1^n)-h(Z_2^n)\\
&\leq \sum_{t=1}^n[h(Y_1(t)|S_1(t))+h(Y_2(t)|S_2(t))-h(Z_1(t))-h(Z_2(t))]
\end{align}
where the second inequality follows from the chain rule and the fact that dropping conditioning does not reduce entropy. Finally, because the circularly symmetric complex Gaussian distribution maximizes conditional differential entropy for a given covariance constraint, we obtain
\begin{equation}
\begin{aligned}
\sum_{i=1}^2\sum_{j=1}^3R_{ij}-2\epsilon \leq& \log_2\left(1+P^{\alpha_{13}}+P^{\alpha_{12}}+\frac{P^{\alpha_{11}}}{1+P^{\alpha_{21}}}\right)\\
&+\log_2\left(1+P^{\alpha_{23}}+P^{\alpha_{21}}+\frac{P^{\alpha_{22}}}{1+P^{\alpha_{12}}}\right)
\end{aligned}
\end{equation}
Due to the condition (\ref{cond}), in the GDoF sense we obtain
\begin{align}
\hat{d}_1+\hat{d}_2\leq (\alpha_{11}+\alpha_{22})-(\alpha_{12}+\alpha_{21})
\end{align}
which is the desired extension, (\ref{ex2_X_cycle}), to the X channel setting of the original bound, (\ref{ex2_IC_cycle}), for the interference channel.
\hfill $\Box$

\end{example}

\bigskip

Now let us consider the proof for the general $K\times K$ $X$ channel. For the individual bounds in the $K$-user interference channel
\begin{align}
d_i\leq \alpha_{ii}~~\forall i\in\{1,2,...,K\},
\end{align}
in its counterpart $X$ channel, the corresponding bound comes from the MAC consisting of all the transmitters and the receiver $i$,
\begin{align}
\sum_{j=1}^KR_{ij}\leq\log_2(1+\sum_{j=1}^KP^{\alpha_{ij}})
\end{align}
According to (\ref{cond}), in the GDoF sense we have
\begin{align}
\hat{d}_i=\sum_{j=1}^Kd_{ij}\leq\alpha_{ii}
\end{align}

\begin{figure}[h]
\begin{center}
 \includegraphics[width= 6 cm]{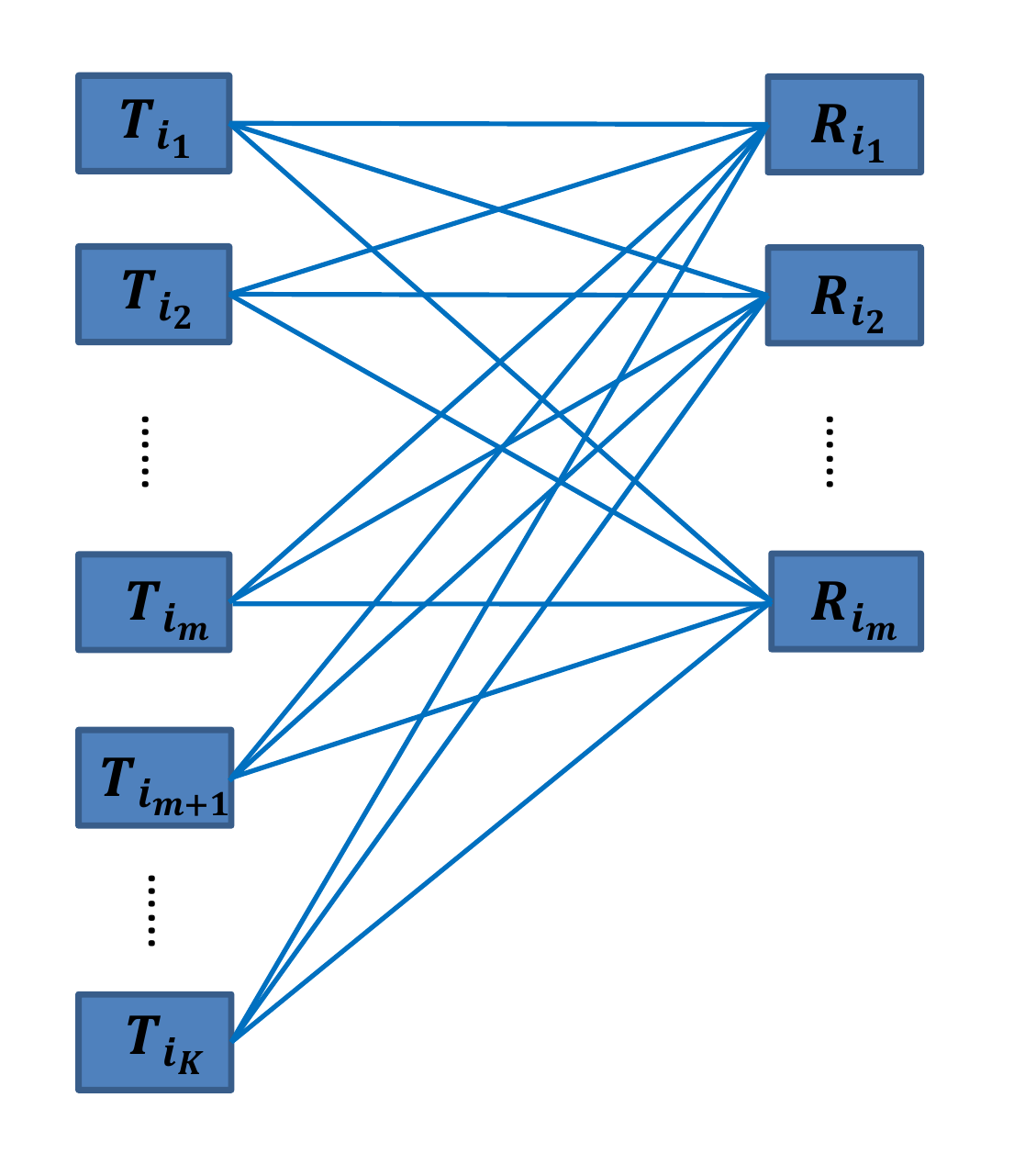}
 \caption{A $K\times m$ $X$ channel ($K\geq m$)}
\label{converse}
\end{center}
\end{figure}

For any cycle bound in the interference channel
\begin{align}
\sum_{j=1}^{m}d_{i_j}\leq \sum_{j=1}^{m}(\alpha_{i_ji_j}-\alpha_{i_{j-1}i_j}),\:\:&&\forall (i_1,...,i_m)\in\Pi_K,~~\forall m\in\{2,3,...,K\},
\end{align}
consider the subnetwork consisting of all the transmitters and the receivers $\{i_1,i_2,...,i_m\}$ as shown in Fig.~\ref{converse}. Eliminate all other receivers and their desired messages, which cannot hurt the rates of the remaining messages. For such a $K\times m$ $X$ channel, define $\mathcal{W}\triangleq\{W_{i_ji_k}\}$, $\mathcal{W}_{i_j}^*\triangleq\{W_{i_ji_1},W_{i_ji_2},...,W_{i_ji_K}\}$, $\mathcal{W}_{i_k}^\dag\triangleq\{W_{i_1i_k},W_{i_2i_k},...,W_{i_mi_k}\}$, and $\mathcal{W}_\mathcal{S}^c\triangleq\mathcal{W}/\mathcal{W}_\mathcal{S}$, where $\forall j\in\{1,2,...,m\}$, $\forall k\in\{1,2,...,K\}$, and $\mathcal{S}$ is any subset of message indices. In words, the sets $\mathcal{W}$, $\mathcal{W}_{i_j}^*$, and $\mathcal{W}_{i_k}^\dag$ represent all the remaining messages delivered in the channel, all the messages intended to receiver $i_j$, and all the messages coming from transmitter $i_k$, respectively, and $\mathcal{W}_\mathcal{S}^c$ is the complement of $\mathcal{W}_\mathcal{S}$ in $\mathcal{W}$. For instance, when $j,k\in\{1,2\}$ and $\mathcal{S}=\{i_1i_1,i_1i_2\}$, then $\mathcal{W}_\mathcal{S}=\{W_{i_1i_1},W_{i_1i_2}\}$ and $\mathcal{W}_\mathcal{S}^c=\{W_{i_2i_1},W_{i_2i_2}\}$. Modulo-$m$ arithmetic is used on the receiver indices, e.g., $i_0=i_m$. Lastly, to complete the setup, define
\begin{align}
S_{i_j}(t)=h_{i_{j-1}i_j}X_{i_j}(t)+Z_{i_{j-1}}(t) ~~ \forall j\in\{1,2,...,m\}
\end{align}

Then for receiver $i_1$, we provide $S_{i_1}^n$, $\mathcal{W}_{i_2i_2}^c/\mathcal{W}_{i_1}^*$ through a genie.  From Fano's inequality, we have
\begin{align}
&n(\sum_{k=1}^KR_{i_1i_k}-\epsilon)\\
&\leq I(\mathcal{W}_{i_1}^*;Y_{i_1}^n,S_{i_1}^n,\mathcal{W}_{i_2i_2}^c/\mathcal{W}_{i_1}^*)\\
&=I(\mathcal{W}_{i_1}^*;Y_{i_1}^n,S_{i_1}^n|\mathcal{W}_{i_{2}i_{2}}^c/\mathcal{W}_{i_1}^*)\label{converse_S1}\\
&=I(\mathcal{W}_{i_1}^*;S_{i_1}^n|\mathcal{W}_{i_{2}i_{2}}^c/\mathcal{W}_{i_1}^*)+I(\mathcal{W}_{i_1}^*;Y_{i_1}^n|S_{i_1}^n,\mathcal{W}_{i_{2}i_{2}}^c/\mathcal{W}_{i_1}^*)\\
&=h(S_{i_1}^n|\mathcal{W}_{i_{2}i_{2}}^c/\mathcal{W}_{i_1}^*)-h(S_{i_1}^n|\mathcal{W}_{i_{2}i_{2}}^c)+h(Y_{i_1}^n|S_{i_1}^n,\mathcal{W}_{i_{2}i_{2}}^c/\mathcal{W}_{i_1}^*)-h(Y_{i_1}^n|S_{i_1}^n,\mathcal{W}_{i_{2}i_{2}}^c)\\
&\leq h(S_{i_1}^n|\mathcal{W}_{i_1}^\dag/W_{i_1i_1})-h(Z_{i_{0}}^n)+h(Y_{i_1}^n|S_{i_1}^n)-h(S_{i_{2}}^n|\mathcal{W}_{i_{2}}^\dag/W_{i_{2}i_{2}})\label{converse_S2}
\end{align}
where (\ref{converse_S1}) follows because all the messages are independent, and in (\ref{converse_S2}) we use the fact that dropping conditioning does not reduce entropy.

Similarly, for other receivers $i_j$, $\forall j\in\{2,3,...,m-1\}$, by providing $S_{i_j}^n$, $\mathcal{W}_{i_{j+1}i_{j+1}}^c/\mathcal{W}_{i_j}^*$ through a genie we have
\begin{align}
n(\sum_{k=1}^KR_{i_ji_k}-\epsilon)\leq h(S_{i_j}^n|\mathcal{W}_{i_j}^\dag/W_{i_ji_j})-h(Z_{i_{j-1}}^n)+h(Y_{i_j}^n|S_{i_j}^n)-h(S_{i_{j+1}}^n|\mathcal{W}_{i_{j+1}}^\dag/W_{i_{j+1}i_{j+1}})
\end{align}
Finally for receiver $i_m$, we can provide $S_{i_m}^n$, $\mathcal{W}_{i_1i_1}^c/\mathcal{W}_{i_m}^*$ through a genie and obtain
\begin{align}
n(\sum_{k=1}^KR_{i_mi_k}-\epsilon)\leq h(S_{i_m}^n|\mathcal{W}_{i_m}^\dag/W_{i_mi_m})-h(Z_{i_{m-1}}^n)+h(Y_{i_m}^n|S_{i_m}^n)-h(S_{i_{1}}^n|\mathcal{W}_{i_{1}}^\dag/W_{i_{1}i_{1}})
\end{align}

Then taking the sum of $n(\sum_{k=1}^KR_{i_ji_k}-\epsilon)$ for all $j\in\{1,2,...,m\}$, we have
\begin{align}
n(\sum_{j=1}^m\sum_{k=1}^KR_{i_ji_k}-m\epsilon)&\leq \sum_{j=1}^m[h(Y_{i_j}^n|S_{i_j}^n)-h(Z_{i_j}^n)]\\
&\leq \sum_{t=1}^n\sum_{j=1}^m[h(Y_{i_j}(t)|S_{i_j}(t))-h(Z_{i_j}(t))]\label{converse_S3}
\end{align}
where (\ref{converse_S3}) follows the chain rule and the fact that dropping conditioning does not reduce entropy. Once again, using the fact that the circularly symmetric complex Gaussian distribution maximizes conditional differential entropy for a given covariance constraint and the condition (\ref{cond}), we can obtain the following desired outer bound in the GDoF sense, through the same set of  manipulations as in Example \ref{ex2},
\begin{align}
\sum_{j=1}^{m}\hat{d}_{i_j} = \sum_{j=1}^m\sum_{k=1}^Kd_{i_ji_k}\leq \sum_{j=1}^{m}(\alpha_{i_ji_j}-\alpha_{i_{j-1}i_j})
\end{align}

\bigskip


Now we can proceed to the last step to prove that under condition (\ref{cond}), the $K$-user interference channel and its counterpart $K\times K$ $X$ channel have the same sum-GDoF. According to Theorem \ref{tin-IC}, for the $K$-user interference channel, under condition (\ref{cond}), to obtain its sum-GDoF $d_{\Sigma,IC}$, we need to solve the following linear programming (LP) problem
\begin{align}
&\max \sum_{i=1}^Kd_i\\
\mathrm{s.t.}~~& 0\leq d_i\leq \alpha_{ii}\:\:&&\forall i\in\{1,2,...,K\}\label{cnstr1}\\
&\sum_{j=1}^{m}d_{i_j}\leq \sum_{j=1}^{m}(\alpha_{i_ji_j}-\alpha_{i_{j-1}i_j}),\:\:&&\forall (i_0,i_1,...,i_m)\in\Pi_K,~~\forall m\in\{2,3,...,K\}\label{constr2}
\end{align}
To get the sum-GDoF of its counterpart $X$ channel $d_{\Sigma,X}$, we consider a similar LP problem. Note for this LP problem, with the objective function $\sum_{i=1}^K\hat{d}_i$, it needs to follow similar constraints to (\ref{cnstr1}) and (\ref{constr2}), in which each $d_i$ is just replaced by $\hat{d}_i$. Thus we have $d_{\Sigma,IC}\geq d_{\Sigma,X}$. Obviously, in any case, the sum-GDoF of the $K$-user interference channel must be less than or equal to that of its counterpart $X$ channel, i.e. $d_{\Sigma,IC}\leq d_{\Sigma,X}$. Therefore, under condition (\ref{cond}), we have established that the $K$-user interference channel and its counterpart $X$ channel have the same sum-GDoF.



\bigskip
\textbf{Proof for the Constant Gap of Sum Capacity}:
Based on the insight gained in the above GDoF study, for the TIN-optimal $K\times K$ $X$ channel, we intend to characterize the sum channel capacity to within a constant gap of no more than $K\log_2[K(K+1)]$ bits. To this end, first recall the achievability proof in \cite{Geng_TIN_opt}. By operating the $K\times K$ $X$ channel as an interference channel, in which each transmitter $i$ sends one independent message $W_i$ to its corresponding receiver $i$ ($\forall i\in\{1,2,...,K\}$), power control and TIN can achieve the following rate tuples $(R_{1,\mathrm{TIN}},R_{2,\mathrm{TIN}},...,R_{K,\mathrm{TIN}})$ satisfying
\begin{align}
R_{i,\mathrm{TIN}}&\leq\alpha_{ii}\log_2P+\log_2(\frac{1}{K})\label{Rate_IB1}~~~~\forall i\in\{1,2,...,K\}\\
\sum_{j=1}^mR_{i_j,\mathrm{TIN}}&=\sum_{j=1}^m[d_{i_j}\log_2P+\log_2(\frac{1}{K})]\nonumber\\
&\leq\sum_{j=1}^m[(\alpha_{i_ji_j}-\alpha_{i_{j-1}i_j})\log_2P+\log_2(\frac{1}{K})],\label{Rate_IB2}
\end{align}
for all cycles $(i_0,i_1,...,i_m)\in\Pi_K$, $\forall m\in\{2,3,...,K\}$.

Next consider the converse. Start with the individual bounds,
\begin{align}
\hat{R}_i&=\sum_{j=1}^KR_{ij}\\
&\leq\log_2(1+\sum_{j=1}^KP^{\alpha_{ij}})\\
&\leq\log_2[(K+1)P^{\alpha_{ii}}]\\
&=\alpha_{ii}\log_2 P+\log_2(K+1)\label{Rate_OB1}
\end{align}

Then for the cycle bounds, from (\ref{converse_S3}), it is easy to obtain
\begin{align}
\sum_{j=1}^m\hat{R}_{i_j}&\leq\sum_{j=1}^m\log_2[\frac{(K+1)P^{\alpha_{i_ji_j}}}{P^{\alpha_{i_{j-1}i_j}}}]\\
&=\sum_{j=1}^m[(\alpha_{i_ji_j}-\alpha_{i_{j-1}i_j})\log_2 P + \log_2(K+1)]\label{Rate_OB2}
\end{align}
for all cycles $(i_0,i_1,...,i_m)\in\Pi_K$, $\forall m\in\{2,3,...,K\}$.

Comparing (\ref{Rate_IB1}) and (\ref{Rate_IB2}) with (\ref{Rate_OB1}) and (\ref{Rate_OB2}), we can characterize the sum channel capacity to within a constant gap of no more than $K\log_2[K(K+1)]$ bits, which is only dependent on the number of users $K$.

\subsection{Proof for Theorem \ref{tin-X-asy}} \label{proofs2}

It is easy to verify that when $M\geq N$, by defining $\hat{d}_i=\sum_{j=1}^Md_{ij}$ ($\forall i\in\{1,2,...,N\}$) and following the same argument as in the proof of  Theorem \ref{tin-X-GDoF}, we can complete the proof. Therefore, hereafter we only consider the case where $\kappa=M<N$.

Without loss of generality, we assume that the two permutations $\Pi^T$ and $\Pi^R$ satisfying the condition (\ref{cond_asy}) are $\Pi^T=\{1,2,...,M\}$ and $\Pi^R=\{1,2,...,N\}$, i.e.,
\begin{align}\label{cond_asy1}
\alpha_{ii}\geq \max_{j:j\neq i}\{\alpha_{ji}\}+\max_{k:k\neq i}\{\alpha_{ik}\}~~\forall i\in\{1,2,...,M\},\forall j\in\{1,2,...,N\},\forall k\in\{1,2,...,M\}
\end{align}


In this case, similar to the proof of Theorem \ref{tin-X-GDoF}, the key step is to show that when (\ref{cond_asy1}) holds, then for each individual bound and cycle bound  in the  $M$-user interference channel consisting of transmitters $\{1,2,...,M\}$ and receivers $\{1,2,...,M\}$,  if each $d_i$ ($\forall i\in\{1,2,...,M\}$) is replaced by $\bar{d}_i=\sum_{j=1}^Nd_{ji}$, the resulting bounds  hold in the $M\times N$ $X$ channel. Then based on the same argument of Theorem \ref{tin-X-GDoF}, we can prove the optimality of TIN for the sum-GDoF of the $M\times N$ $X$ channel where $M<N$.

For the individual bounds, consider the degraded broadcast channel (BC) comprised of  transmitter $i$ ($\forall i \in\{1,2,...,M\}$) and all the receivers, eliminating all other transmitters and their messages. Since (\ref{cond_asy1}) is satisfied, receiver $i$ is the strongest receiver, and can decode all the messages from transmitter $i$. Thus in the GDoF sense we have
\begin{align}
\bar{d}_i=\sum_{j=1}^Nd_{ji}\leq \alpha_{ii}~~\forall i\in\{1,2,...,M\}
\end{align}

\begin{figure}[h]
\begin{center}
 \includegraphics[width= 8 cm]{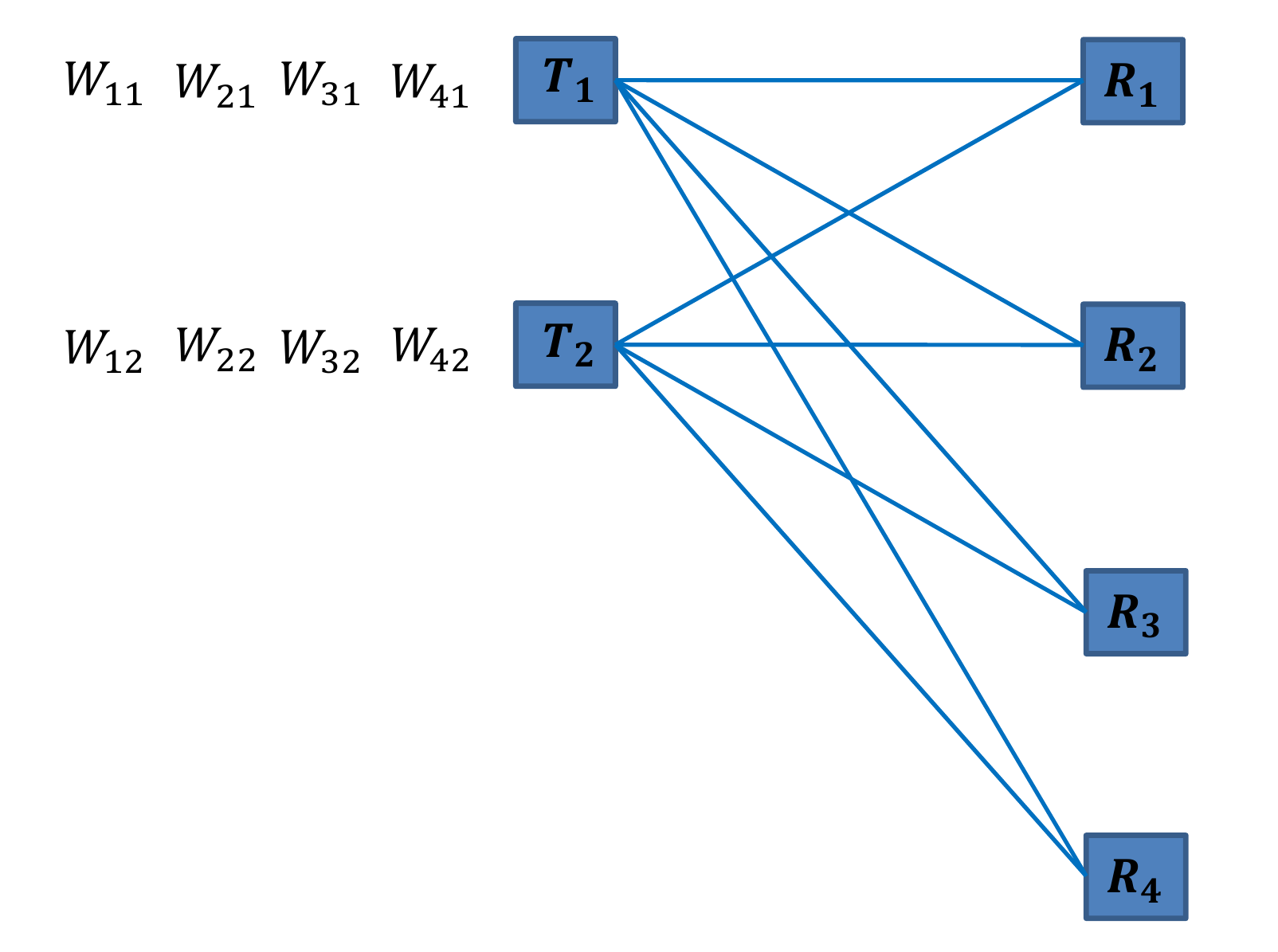}
 \caption{A $2\times 4$ $X$ channel}
\label{2-4user_X}
\end{center}
\end{figure}

Now the only task left is to prove that in the $X$ setting, by replacing $d_i$ with $\bar{d}_i$, all the cycle bounds still hold. Before exploring the proof details, let's see an intuitive sketch of proof first for a $2\times 4$ $X$ channel as illustrated in Fig.~\ref{2-4user_X}. For this $X$ channel, when (\ref{cond_asy1}) is satisfied, we intend to prove
\begin{align*}
\bar{d}_1+\bar{d}_2\leq (\alpha_{11}+\alpha_{22})-(\alpha_{12}+\alpha_{21})
\end{align*}

\emph{An intuitive sketch of proof for the above cycle bound}:
In this $2\times 4$ $X$ channel, we assume $\alpha_{31}\geq \alpha_{41}$ and $\alpha_{42}\geq \alpha_{32}$. The proof for all the other cases follows similarly. Define the message set $\mathcal{\widetilde{W}}\triangleq\{W_{ki}\}$, $\forall i\in\{1,2\}$, $\forall k\in\{1,2,3,4\}$. Also define
\begin{align*}
S_1(t)&=h_{21}X_1(t)+Z_2(t)\\
S_2(t)&=h_{12}X_2(t)+Z_1(t)
\end{align*}

Start with receiver $1$, we have
\begin{align}
&n(R_{11}+R_{12}-\epsilon)\\
&\leq I(W_{11},W_{12};Y_{1}^n,S_1^n|W_{21})\\
&=h(Y_{1}^n,S_1^n|W_{21})-h(Y_{1}^n,S_1^n|W_{21},W_{11},W_{12})\\
&=h(S_1^n|W_{21})+h(Y_{1}^n|S_1^n,W_{21})-h(Y_{1}^n|W_{21},W_{11},W_{12})-h(S_1^n|Y_{1}^n,W_{21},W_{11},W_{12})\\
&\leq h(S_1^n|W_{21})+h(Y_{1}^n|S_1^n)-h(Y_{1}^n|W_{21},W_{11},W_{12})-h(Z_2^n) \\
&\leq h(S_1^n|W_{21})+h(Y_{1}^n|S_1^n)-h(Y_{3}^n|\mathcal{\widetilde{W}}_{\{31,41\}}^c)-h(S_2^n|W_{12})-h(Z_2^n)-n~o(\log(P)) \label{PS_R1}
\end{align}
where $\mathcal{\widetilde{W}}_{\{31,41\}}^c$ denotes the complement of $\{W_{31},W_{41}\}$ in $\mathcal{\widetilde{W}}$. The last inequality is the key step of the proof. Intuitively, it is due to the fact that out of the $\alpha_{11}\log(P)$ bit levels of $Y_1$ that are above the noise floor, $S_2$ is contained in the lowest $\alpha_{12}\log(P)$ bit levels of $Y_1$, whereas only the top $\alpha_{31}\log(P)$ bit levels are seen by receiver $3$. Since $\alpha_{11}\geq \alpha_{12}+\alpha_{31}$, these bit levels do not overlap, i.e., they can be recovered from $Y_1$ within a bounded entropy gap.

Then consider the degraded BC comprised of the transmitter $1$ and the receivers $3$ and $4$. Since $\alpha_{31}\geq \alpha_{41}$, we have
\begin{align}
n(R_{31}+R_{41}-\epsilon)\leq & I(W_{31},W_{41};Y_3^n|\mathcal{\widetilde{W}}_{\{31,41\}}^c)\\
=&h(Y_3^n|\mathcal{\widetilde{W}}_{\{31,41\}}^c)-h(Z_3^n)\label{PS_BC}
\end{align}

Adding (\ref{PS_R1}) and (\ref{PS_BC}), we obtain
\begin{align}
&n(R_{11}+R_{12}+R_{31}+R_{41}-\epsilon)\\
&\leq  h(S_1^n|W_{21})+h(Y_{1}^n|S_1^n)-h(S_2^n|W_{12})-n~o(\log(P)) \label{PS_R1_2}
\end{align}

Similarly, we have
\begin{align}
&n(R_{21}+R_{22}+R_{32}+R_{42}-\epsilon)\\
&\leq  h(S_2^n|W_{12})+h(Y_{2}^n|S_2^n)-h(S_1^n|W_{21})-n~o(\log(P)) \label{PS_R2}
\end{align}

Finally, through adding (\ref{PS_R1_2}) and (\ref{PS_R2}) together and some other manipulations, we can obtain the desired outer bound,
\begin{align*}
n(R_{\Sigma}-\epsilon) &\leq h(Y_{1}^n|S_1^n)+ h(Y_{2}^n|S_2^n) - n~o(\log(P))\\
\Rightarrow\bar{d}_1+\bar{d}_2 &\leq (\alpha_{11}+\alpha_{22})-(\alpha_{12}+\alpha_{21})
\end{align*}

\bigskip

In the following, in order to make the intuitive justification of the key step (\ref{PS_R1}) rigorous, we take a \emph{deterministic} approach \cite{Huang_GDoF_X, Niesen_Maddah, Gamal_Deterministic, ADT}. We first show that the sum capacity of the original complex Gaussian $X$ channel is upper bounded by that of one suitably-chosen deterministic channel up to a constant gap. Then by upper bounding that deterministic channel, we obtain the desired converse of the original Gaussian channel as well.

Recall the original complex Gaussian $X$ channel. Denote
\begin{align*}
&X_k(t)=X_k^R(t)+jX_k^I(t)\\
&h_{ik}=\sqrt{P^{\alpha_{ik}}}e^{j\theta_{ik}}=h_{ik}^R+jh_{ik}^I
\end{align*}
The input-output relationship can be written as
\begin{align}
Y_i(t)=&\sum_{k=1}^M h_{ik}X_{k}(t)+Z_i(t)\\
=&\sum_{k=1}^M\Big[\big(h_{ik}^RX_k^R(t) - h_{ik}^IX_k^I(t) \big) + j\big( h_{ik}^IX_k^R(t)+ h_{ik}^RX_k^I(t)\big)\Big]+Z_i(t),  ~~\forall i\in\{1,2,...,N\}\label{Gaussian_model}
\end{align}
where $ E[|X_i(t)|^2]\leq 1$ and $Z_i(t) \sim \mathcal{CN}(0,1)$. By scaling the output, we may set
\begin{align*}
E[|X_i(t)|^2]\leq 2, ~Z_i(t) \sim \mathcal{CN}(0,2).
\end{align*}

In this paper, we consider the following deterministic model,
\begin{equation}
\begin{aligned}\label{deter_model}
\hat{Y}_i(t)=&\sum_{k=1}^{M}\Big[\big(\lfloor \mathrm{sign}(\bar{X}_k^R(t))h_{ik}^R\sum_{b=1}^{m_{ik}^R}\bar{X}_{k,b}^R(t)2^{-b}\rfloor -\lfloor \mathrm{sign}(\bar{X}_k^I(t))h_{ik}^I\sum_{b=1}^{m_{ik}^I}\bar{X}_{k,b}^I(t)2^{-b}\rfloor\big)\\
&+j\big(\lfloor \mathrm{sign}(\bar{X}_k^R(t))h_{ik}^I\sum_{b=1}^{m_{ik}^I}\bar{X}_{k,b}^R(t)2^{-b}\rfloor + \lfloor \mathrm{sign}(\bar{X}_k^I(t))h_{ik}^R\sum_{b=1}^{m_{ik}^R}\bar{X}_{k,b}^I(t)2^{-b}\rfloor\big)\Big], ~~\forall i\in\{1,2,...,N\}
\end{aligned}
\end{equation}
where $\lfloor x \rfloor$ is the truncation function \emph{which maps $x$ to its integer part}, $m_{ik}^R\triangleq \lfloor \log_2 |h_{ik}^R|\rfloor$, $m_{ik}^I\triangleq \lfloor \log_2 |h_{ik}^I|\rfloor$, the real and imaginary parts of the input signal $\bar{X}_i(t)=\bar{X}_i^R(t)+j\bar{X}_i^I(t)$ both satisfy the unit \emph{peak} power constraint, and $\bar{X}_{i,b}^R(t)$ ($\bar{X}_{i,b}^I(t)$) is the $b$-th bit in the fractional part of $|\bar{X}_i^R(t)|$ ($|\bar{X}_{i,b}^I(t)|$) in the binary expansion\footnote{We can write the real-valued signal $|\bar{X}_i^R|$ ($|\bar{X}_i^R|\leq 1$) in terms of its binary expansion as \begin{align*} |\bar{X}_i^R|=\sum_{b=1}^{\infty}\bar{X}_{i,b}^R2^{-b}=0.\bar{X}_{i,1}^R\bar{X}_{i,2}^R\bar{X}_{i,3}^R... \end{align*}}. For notation brevity, we call the model in (\ref{deter_model}) the \emph{truncated deterministic model}. The following lemma shows that the sum capacity of the Gaussian $X$ channel in (\ref{Gaussian_model}) is upper bounded by that of the truncated deterministic model in (\ref{deter_model}) up to a constant gap.

\begin{lemma}\label{lemma_const}
The sum capacity of the complex Gaussian $X$ channel is upper bounded by the sum capacity of its corresponding truncated deterministic channel up to a constant gap.
\end{lemma}

The proof for the above lemma follows \cite{Bresler_deterministic} and is relegated to  Appendix \ref{appendix-lemma}.

\bigskip
Now define $m_{ij}\triangleq \lfloor \frac{1}{2}\log_2 P^{\alpha_{ij}}\rfloor$. Since $P>1$ and $\alpha_{ii}\geq \alpha_{ij}+\alpha_{ki}$,  $\forall i\notin \{j,k\}$, we have
\begin{align}
&\lfloor \frac{\alpha_{ii}}{2}\log_2 P \rfloor \geq \lfloor \frac{(\alpha_{ij}+\alpha_{ki})}{2}\log_2 P\rfloor\\
\Rightarrow & \lfloor \frac{\alpha_{ii}}{2}\log_2 P \rfloor \geq \lfloor \frac{\alpha_{ij}}{2} \log_2 P \rfloor +\lfloor \frac{\alpha_{ki}}{2}\log_2 P\rfloor\\
\Rightarrow &m_{ii}\geq m_{ij}+m_{ki}~~\forall i,j,k,~~i\notin \{j,k\}\end{align}

In order to convey the key ingredients of the proof more clearly, next we give an example for the \emph{real} Gaussian $2\times 4$ $X$ channel, and then generalize the proof to the \emph{complex} Gaussian $M\times N$ ($M<N$) $X$ channel.

\begin{example}\label{comx-ex}
Consider the real Gaussian $X$ channel with $2$ transmitters and $4$ receivers, where (\ref{cond_asy1}) is satisfied. In this example we still assume $\alpha_{31}\geq \alpha_{41}$ and $\alpha_{42}\geq \alpha_{32}$. As previously mentioned, the proof for all the other cases follows similarly. Also define the message set $\mathcal{\widetilde{W}}\triangleq\{W_{ki}\}$, $\forall i\in\{1,2\}$, $\forall k\in\{1,2,3,4\}$.

Recall that for this $2\times 4$ $X$ channel, we intend to prove the following cycle bound
\begin{align*}
\bar{d}_1+\bar{d}_2\leq \frac{1}{2}[(\alpha_{11}+\alpha_{22})-(\alpha_{12}+\alpha_{21})]
\end{align*}
where the factor $\frac{1}{2}$ is due to the fact that the Gaussian $X$ channel is real-valued.

We start with the corresponding truncated deterministic model. Define
\begin{align}
\hat{S}_1(t)&= \lfloor \mathrm{sign}(\bar{X}_1(t))h_{21}\sum_{b=1}^{m_{21}}\bar{X}_{1,b}(t)2^{-b}\rfloor \\
\hat{S}_2(t)&= \lfloor \mathrm{sign}(\bar{X}_2(t))h_{12}\sum_{b=1}^{m_{12}}\bar{X}_{2,b}(t)2^{-b} \rfloor.
\end{align}



Also define
\begin{align}
\bar{X}_{31,S}(t)=\mathrm{sign}(\bar{X}_1(t))\sum_{b=1}^{m_{31}}\bar{X}_{1,b}(t)2^{-b}
\end{align}

Thus the output of receiver $1$ can be written as
\begin{align}
\hat{Y}_1(t)=&\lfloor \mathrm{sign}(\bar{X}_1(t))h_{11} \sum_{b=1}^{m_{11}}\bar{X}_{1,b}(t)2^{-b}\rfloor + \lfloor \mathrm{sign}(\bar{X}_2(t))h_{12}\sum_{b=1}^{m_{12}}\bar{X}_{2,b}(t)2^{-b} \rfloor\\
=&\lfloor \mathrm{sign}(\bar{X}_1(t))h_{11} \sum_{b=1}^{m_{31}}\bar{X}_{1,b}(t)2^{-b}\rfloor+ \lfloor  \mathrm{sign}(\bar{X}_1(t))h_{11} \sum_{b=m_{31}+1}^{m_{11}}\bar{X}_{1,b}(t)2^{-b} \rfloor + \hat{S}_2(t)+\hat{C}_1(t)\\
=&\underbrace{\lfloor h_{11} \bar{X}_{31,S}(t)\rfloor}_{\hat{Y}_{1,u}(t)} + \underbrace{\lfloor  \mathrm{sign}(\bar{X}_1(t))h_{11} \sum_{b=m_{31}+1}^{m_{11}}\bar{X}_{1,b}(t)2^{-b} \rfloor + \hat{S}_2(t)}_{\hat{Y}_{1,l}(t)}+\hat{C}_1(t)
\end{align}
where $\hat{C}_1(t)$ may take a value from $\{-1,0,1\}$.

For receiver $1$, we have
\begin{align}
&n(R_{11}+R_{12}-\epsilon)\\
&\leq I(W_{11},W_{12};\hat{Y}_{1,u}^n,\hat{Y}_{1,l}^n,\hat{C}_1^n,\hat{S}_1^n|W_{21})\\
&=H(\hat{Y}_{1,u}^n,\hat{Y}_{1,l}^n,\hat{C}_1^n,\hat{S}_1^n|W_{21})-H(\hat{Y}_{1,u}^n,\hat{Y}_{1,l}^n,\hat{C}_1^n,\hat{S}_1^n|W_{21},W_{11},W_{12})\\
&=H(\hat{S}_1^n|W_{21})+H(\hat{Y}_{1,u}^n,\hat{Y}_{1,l}^n,\hat{C}_1^n|\hat{S}_1^n,W_{21})-H(\hat{Y}_{1,u}^n,\hat{Y}_{1,l}^n,\hat{C}_1^n|W_{21},W_{11},W_{12})\\
&~~~-H(S_1^n|\hat{Y}_{1,u}^n,\hat{Y}_{1,l}^n,\hat{C}_1^n,W_{21},W_{11},W_{12})\\
&\leq H(\hat{S}_1^n|W_{21})+H(\hat{Y}_{1,u}^n,\hat{Y}_{1,l}^n,\hat{C}_1^n,|\hat{S}_1^n)-H(\hat{Y}_{1,u}^n,\hat{Y}_{1,l}^n,\hat{C}_1^n,|W_{21},W_{11},W_{12}) \label{R1}
\end{align}
where (\ref{R1}) follows that dropping conditioning does not reduce entropy. Now consider the last term in (\ref{R1}),
\begin{align}
&H(\hat{Y}_{1,u}^n,\hat{Y}_{1,l}^n,\hat{C}_1^n|W_{21},W_{11},W_{12})\\
&=H(\hat{Y}_{1,u}^n|W_{21},W_{11},W_{12})+H(\hat{Y}_{1,l}^n,\hat{C}_1^n|\hat{Y}_{1,u}^n,W_{21},W_{11},W_{12}) \label{T1}\\
&=H(\bar{X}_{31,S}^n|W_{21},W_{11},W_{12})+H(\hat{Y}_{1,l}^n,\hat{C}_1^n|\hat{Y}_{1,u}^n,W_{21},W_{11},W_{12}) \label{T2}\\
&\geq H(\bar{X}_{31,S}^n|\mathcal{\widetilde{W}}_{\{31,41\}}^c)+H(\hat{S}_2^n|\hat{Y}_{1,u}^n,W_{21},W_{11},W_{12})\\
&= H(\hat{Y}_3^n|\mathcal{\widetilde{W}}_{\{31,41\}}^c)+ H(\hat{S}_2^n|W_{12}) \label{T3}
\end{align}
where (\ref{T2}) holds since the function $f: \bar{X}_{{31},S} \rightarrow \hat{Y}_{1,u}$ is bijective, and (\ref{T3}) follows that conditioning on the messages $\mathcal{\widetilde{W}}_{\{31,41\}}^c$, the function $f: \bar{X}_{31,S} \rightarrow \hat{Y}_{3}$ is bijective.


Plugging (\ref{T3}) into (\ref{R1}), we have
\begin{align}
&n(R_{11}+R_{12}-\epsilon)\leq H(\hat{S}_1^n|W_{21})+H(\hat{Y}_{1,u}^n,\hat{Y}_{1,l}^n,\hat{C}_1^n|\hat{S}_1^n)-H(\hat{S}_2^n|W_{12})-H(\hat{Y}_3^n|\mathcal{\widetilde{W}}_{\{31,41\}}^c) \label{R1_1}
\end{align}

Then consider the degraded BC comprised of the transmitter $1$ and the receivers $3$ and $4$. Since $m_{31}\geq m_{41}$, we have
\begin{align}
n(R_{31}+R_{41}-\epsilon)
&\leq I(W_{31},W_{41};\hat{Y}_3^n|\mathcal{\widetilde{W}}_{\{31,41\}}^c)\\
&=H(\hat{Y}_3^n|\mathcal{\widetilde{W}}_{\{31,41\}}^c)\label{BC}
\end{align}

Combining (\ref{R1_1}) and (\ref{BC}), we obtain
\begin{align}
&n(R_{11}+R_{12}+R_{31}+R_{41}-\epsilon)\\
&\leq  H(\hat{S}_1^n|W_{21})+H(\hat{Y}_{1,u}^n,\hat{Y}_{1,l}^n,\hat{C}_1^n|\hat{S}_1^n)-H(\hat{S}_2^n|W_{12}) \label{R1_2}
\end{align}

Similarly, by considering receiver $2$ and the degraded BC comprised of the transmitter $2$ and the receivers $3$ and $4$, we obtain
\begin{align}
&n(R_{21}+R_{22}+R_{32}+R_{42}-\epsilon)\\
&\leq  H(\hat{S}_2^n|W_{12})+H(\hat{Y}_{2,u}^n,\hat{Y}_{2,l}^n,\hat{C}_2^n|\hat{S}_2^n)-H(\hat{S}_1^n|W_{21}) \label{R2}
\end{align}

Adding (\ref{R1_2}) and (\ref{R2}), the sum capacity of this truncated deterministic $2\times 4$ $X$ channel is upper bounded by
\begin{align}
&n(R_{\Sigma,D}-\epsilon)\\
&\leq H(\hat{Y}_{1,u}^n,\hat{Y}_{1,l}^n,\hat{C}_1^n|\hat{S}_1^n)+H(\hat{Y}_{2,u}^n,\hat{Y}_{2,l}^n,\hat{C}_2^n|\hat{S}_2^n)\\
&\leq \sum_{t=1}^n[H(\hat{Y}_{1,u}(t)|\hat{S}_1(t))+H(\hat{Y}_{1,l}(t)|\hat{S}_1(t))+H(\hat{C}_1(t))+H(\hat{Y}_{2,u}(t)|\hat{S}_2(t))+H(\hat{Y}_{2,l}(t)|\hat{S}_2(t))+H(\hat{C}_2(t))]
\end{align}
where the last inequality follows from the chain rule and the fact that dropping conditioning does not reduce entropy.

Then for the term $H(\hat{Y}_{1,u}(t)|\hat{S}_1(t))+H(\hat{Y}_{1,l}(t)|\hat{S}_1(t))$, we consider two cases,
\begin{itemize}
\item $m_{21}\geq m_{31}$:
\begin{align}
H(\hat{Y}_{1,u}(t)|\hat{S}_1(t))+H(\hat{Y}_{1,l}(t)|\hat{S}_1(t))&\leq 0 + (m_{11}-m_{21})+\mathrm{constant} \label{Case1}\\
&=(m_{11}-m_{21})+\mathrm{constant}
\end{align}
where (\ref{Case1}) follows that conditioning on $\hat{S}_1$, out of the received signal $\hat{Y}_{1,l}$, both the signals from transmitter $1$ and $2$ have at most $m_{11}-m_{21}$ bit-levels, and the sum of two such signals can only induce a loss of constant bits due to carry-overs.


\item{$m_{21}< m_{31}$}: Similarly, we have
\begin{align}
H(\hat{Y}_{1,u}(t)|\hat{S}_1(t))+H(\hat{Y}_{1,l}(t)|\hat{S}_1(t))&\leq (m_{31}-m_{21}) + (m_{11}-m_{31})+\mathrm{constant}\\
&=(m_{11}-m_{21})+\mathrm{constant}
\end{align}
\end{itemize}
Due to symmetry, we always have
\begin{align}
H(\hat{Y}_{2,u}(t)|\hat{S}_2(t))+H(\hat{Y}_{2,l}(t)|\hat{S}_2(t))&\leq (m_{22}-m_{12})+\mathrm{constant}
\end{align}

Therefore,
\begin{align}
n(R_{\Sigma,D}-\epsilon)\leq\sum_{t=1}^n[(m_{11}-m_{21})+(m_{22}-m_{12})+\mathrm{constant}]
\end{align}

According to Lemma \ref{lemma_const}, for the sum capacity of the original Gaussian $X$ channel $R_{\Sigma,G}$, we have
\begin{align}
R_{\Sigma,G}&\leq R_{\Sigma,D} + \mathrm{constant}\\
&\leq (m_{11}-m_{21})+(m_{22}-m_{12})+ \mathrm{constant}\\
&\leq \frac{1}{2}[(\alpha_{11}-\alpha_{21})+(\alpha_{22}-\alpha_{12})]\log_2 P + \mathrm{constant}
\end{align}
Finally, we obtain the desired GDoF cycle bound,
\begin{align}
\bar{d}_1+\bar{d}_2\leq \frac{1}{2}[(\alpha_{11}-\alpha_{21})+(\alpha_{22}-\alpha_{12})].
\end{align}

\hfill $\Box$
\end{example}

\begin{figure}[h]
\begin{center}
 \includegraphics[width= 8 cm]{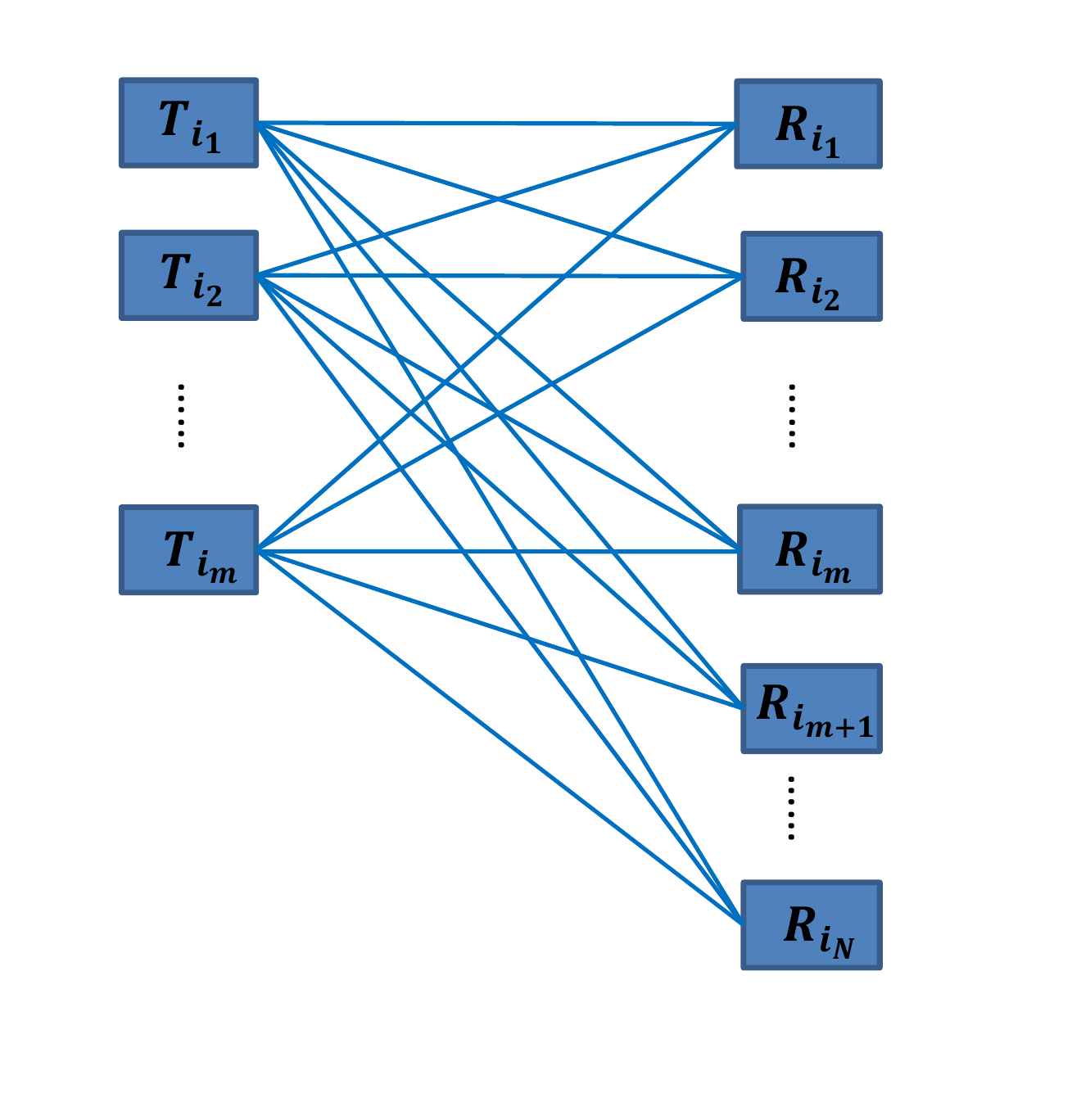}
 \caption{An $m\times N$ $X$ channel ($m < N$)}
\label{M_N-X}
\end{center}
\end{figure}

Now equipped with the bounding techniques in the above example, we can extend the proof to the general complex Gaussian $M\times N$ ($M<N$) $X$ channels. To obtain an arbitrary desired cycle bound
\begin{align}
\sum_{j=1}^{m}\bar{d}_{i_j}\leq \sum_{j=1}^{m}(\alpha_{i_ji_j}-\alpha_{i_{j-1}i_j}),\:\:&&\forall (i_0,i_1,...,i_m)\in\Pi_M,~~\forall m\in\{2,3,...,M\},
\end{align}
consider the subnetwork consisting of all the receivers and the transmitters $\{i_1,i_2,...,i_m\}$ in Fig.~\ref{M_N-X}, eliminating all other transmitters and their messages. First define the message set $\mathcal{\widetilde{W}}\triangleq\{W_{i_ki_j}\}$, $\forall j\in\{1,2,...,m\}$, $\forall k\in\{1,2,...,N\}$. Also define
$\mathcal{W}\triangleq\{W_{i_ki_j}\}$, $\mathcal{W}_{i_k}^*\triangleq\{W_{i_ki_1},W_{i_ki_2},...,W_{i_ki_m}\}$, and $\mathcal{W}_{i_j'}\triangleq\{W_{i_{m+1}i_j},W_{i_{m+2}i_j},...,W_{i_Ni_j}\}$, $\forall j,k\in\{1,2,...,m\}$. Similarly, $\mathcal{W}_\mathcal{S}^c$ denotes $\mathcal{W}/\mathcal{W}_\mathcal{S}$, where $\mathcal{S}$ is a subset of message indices.

\begin{figure}[h]
\begin{center}
 \includegraphics[width= 11 cm]{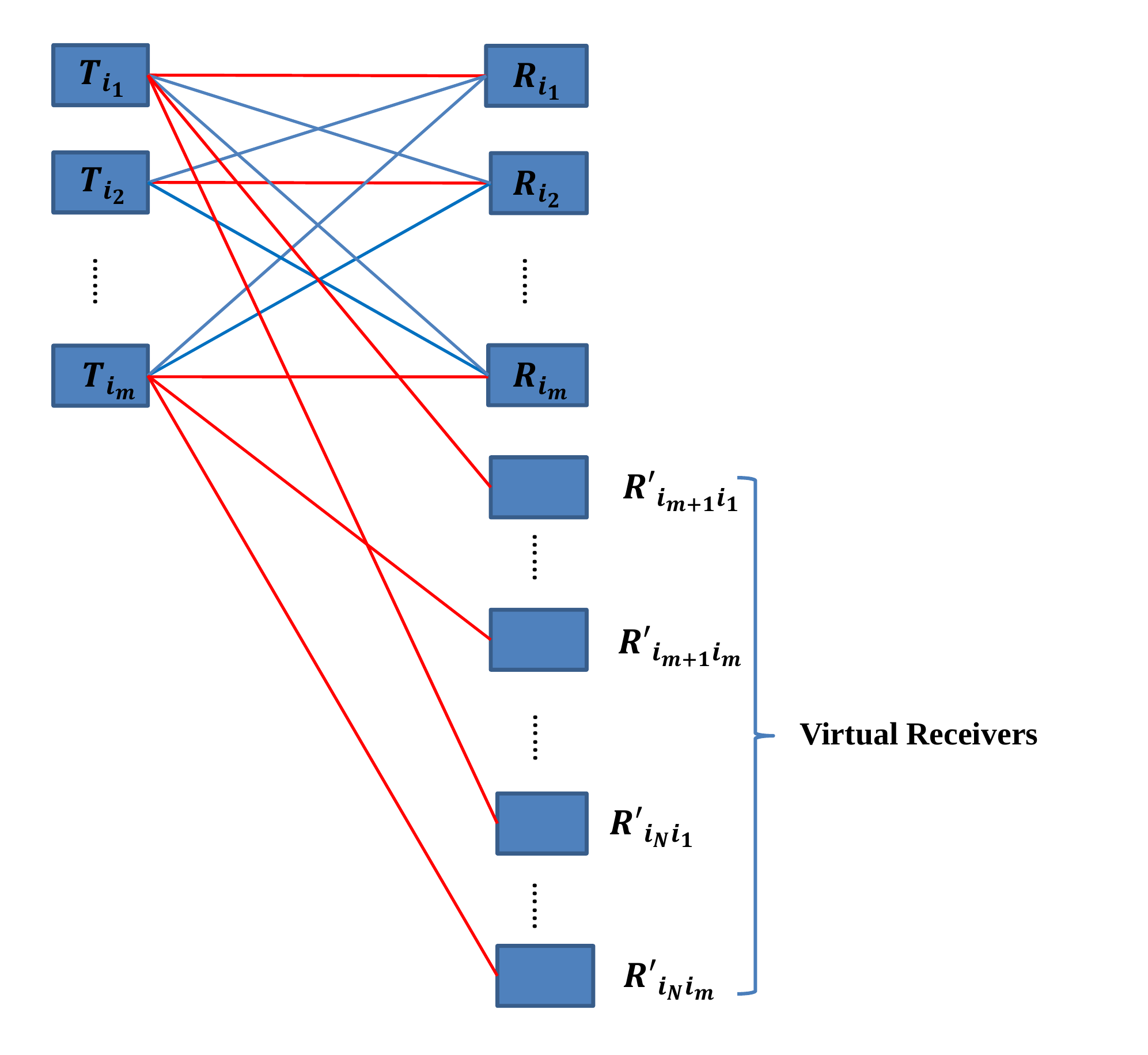}
 \caption{The constructed channel which upper-bounds the sum channel capacity of the $m\times N$ $X$ channel in Fig.~\ref{M_N-X}. The red links are real-valued by rotating the phase of the received signal at the corresponding receivers.}
\label{M_N-X-mod}
\end{center}
\end{figure}

To simplify the proof, we construct the following channel as shown in Fig.~\ref{M_N-X-mod}, which upper bounds the sum channel capacity of the original complex Gaussian $X$ channel in Fig.~\ref{M_N-X}:
\begin{itemize}
\item Step $1$: We start with an $m\times m$ $X$ channel with channel coefficients $h_{i_ki_j}$, $\forall k,j\in\{1,2,...,m\}$.

\item Step $2$: For each transmitter $i_j$, $\forall j\in\{1,2,...,m\}$, we create another $N-m$ virtual receivers. The virtual receiver $R_{i_ki_j}'$, $\forall k\in\{m+1,m+2,...,N\}$, only connects to the transmitter $i_j$ with the channel coefficient $h_{i_ki_j}$ and desires the message $W_{i_ki_j}$ from transmitter $i_j$. Note now there are $m\times N$ messages totally in the network.

\item Step $3$: For the receiver $i_k$, $\forall k\in\{1,2,...,m\}$, it rotates the channel output appropriately to make $h_{i_ki_k}$ real-valued. Similarly, for the virtual receiver $R_{i_ki_j}'$, $\forall j\in\{1,2,...,m\}$, $\forall k\in\{m+1,m+2,...,N\}$, it rotates the channel output to make its only connected link real-valued.

\item Step $4$: The input signal $X_{i_j}(t)$ satisfies the power constraint $E[|X_{i_j}(t)|^2]\leq 2$, $\forall j\in\{1,2,...,m\}$, and the AWGN seen at all the receivers are independent and with zero mean and variance $2$.
\end{itemize}

For the constructed channel in Fig.~\ref{M_N-X-mod}, consider its corresponding truncated deterministic model. For receivers $i_j$, $\forall j\in\{1,2,...,m\}$, the channel output can be written in the following matrix form,
\begin{align}
\hat{Y}_{i_j}(t)=&\left(
            \begin{array}{c}
              \lfloor \mathrm{sign}(\bar{X}_{i_j}^R(t))h_{i_ji_j}^R\sum_{b=1}^{m_{i_ji_j}^R}\bar{X}_{i_j,b}^R(t)2^{-b} \rfloor \\
              \lfloor \mathrm{sign}(\bar{X}_{i_j}^I(t))h_{i_ji_j}^R\sum_{b=1}^{m_{i_ji_j}^R}\bar{X}_{i_j,b}^I(t)2^{-b} \rfloor \\
            \end{array}
          \right)\\
          &+ \sum_{k=1,k\neq j}^m
          \left(
            \begin{array}{c}
              \lfloor \mathrm{sign}(\bar{X}_{i_k}^R(t))h_{i_ji_k}^R\sum_{b=1}^{m_{i_ji_k}^R}\bar{X}_{i_k,b}^R(t)2^{-b} \rfloor - \lfloor \mathrm{sign}(\bar{X}_{i_k}^I(t))h_{i_ji_k}^I\sum_{b=1}^{m_{i_ji_k}^I}\bar{X}_{i_k,b}^I(t)2^{-b} \rfloor  \\
              \lfloor \mathrm{sign}(\bar{X}_{i_k}^R(t))h_{i_ji_k}^I\sum_{b=1}^{m_{i_ji_k}^I}\bar{X}_{i_k,b}^R(t)2^{-b} \rfloor + \lfloor \mathrm{sign}(\bar{X}_{i_k}^I(t))h_{i_ji_k}^R\sum_{b=1}^{m_{i_ji_k}^R}\bar{X}_{i_k,b}^I(t)2^{-b} \rfloor\\
            \end{array}
          \right)
          \end{align}
While for the virtual receivers, the channel output is
\begin{align}
\hat{Y}_{i_ki_j}(t)=&
          \left(
            \begin{array}{c}
              \lfloor \mathrm{sign}(\bar{X}_{i_j}^R(t))h_{i_ki_j}^R\sum_{b=1}^{m_{i_ki_j}^R}\bar{X}_{i_j,b}^R(t)2^{-b} \rfloor\\
              \lfloor \mathrm{sign}(\bar{X}_{i_j}^I(t))h_{i_ki_j}^R\sum_{b=1}^{m_{i_ki_j}^R}\bar{X}_{i_j,b}^I(t)2^{-b} \rfloor \\
            \end{array}
          \right) ~~ \forall j\in\{1,2,...,m\},~\forall k\in\{m+1,m+2,...,N\}.
\end{align}

Next we present a  lemma that will be useful later. The proof is presented in Appendix \ref{appendix-lemma2}.
\begin{lemma}\label{complex_map}

For $\forall j\in\{1,2,...,m\}$, define
\begin{align}
\hat{S}_{i_j}(t)=&\left(
                 \begin{array}{c}
                   \mathrm{sign}(\bar{X}_{i_j}^R(t))\sum_{b=1}^{\max\{m_{i_{j-1}i_j}^R,m_{i_{j-1}i_j}^I\}}\bar{X}_{i_j,b}^R(t)2^{-b} \\
                   \mathrm{sign}(\bar{X}_{i_j}^I(t))\sum_{b=1}^{\max\{m_{i_{j-1}i_j}^R,m_{i_{j-1}i_j}^I\}}\bar{X}_{i_j,b}^I(t)2^{-b} \\
                 \end{array}
               \right), \\
\hat{S}_{i_j}'(t)=&
          \left(
            \begin{array}{c}
              \lfloor \mathrm{sign}(\bar{X}_{i_j}^R(t))h_{i_{j-1}i_j}^R\sum_{b=1}^{m_{i_{j-1}i_j}^R}\bar{X}_{i_j,b}^R(t)2^{-b} \rfloor - \lfloor \mathrm{sign}(\bar{X}_{i_j}^I(t))h_{i_{j-1}i_j}^I\sum_{b=1}^{m_{i_{j-1}i_j}^I}\bar{X}_{i_j,b}^I(t)2^{-b} \rfloor \\
              \lfloor \mathrm{sign}(\bar{X}_{i_j}^R(t))h_{i_{j-1}i_j}^I\sum_{b=1}^{m_{i_{j-1}i_j}^I}\bar{X}_{i_j,b}^R(t)2^{-b} \rfloor + \lfloor \mathrm{sign}(\bar{X}_{i_j}^I(t))h_{i_{j-1}i_j}^R\sum_{b=1}^{m_{i_{j-1}i_j}^R}\bar{X}_{i_j,b}^I(t)2^{-b} \rfloor \\
            \end{array}
          \right),
\end{align}
where the modulo-$m$ arithmetic is implicitly used on the user indices, e.g., for $i_0=i_m$. Then $f:\hat{S}_{i_j}(t)\rightarrow\hat{S}'_{i_j}(t)$ is bijective.
\end{lemma}


\bigskip

For receiver $i_1$, its output can be rewritten as
\begin{align}
\hat{Y}_{i_1}(t)=&\left(
   \begin{array}{c}
     \hat{Y}_{i_1}^R(t) \\
     \hat{Y}_{i_1}^I(t) \\
   \end{array}
 \right)\\
 =& \underbrace{
          \left(
            \begin{array}{c}
              \lfloor \mathrm{sign}(\bar{X}_{i_1}^R(t))h_{i_1i_1}^R\sum_{b=1}^{m_{i_1^*i_1}^R}\bar{X}_{i_1,b}^R(t)2^{-b} \rfloor \\
              \lfloor \mathrm{sign}(\bar{X}_{i_1}^I(t))h_{i_1i_1}^R\sum_{b=1}^{m_{i_1^*i_1}^R}\bar{X}_{i_1,b}^I(t)2^{-b} \rfloor \\
            \end{array}
          \right)}_{\hat{Y}_{i_1,u}(t)}\\
          &
          +\left(
            \begin{array}{c}
              \lfloor \mathrm{sign}(\bar{X}_{i_1}^R(t))h_{i_1i_1}^R\sum_{b=m_{i_1^*i_1}^R+1}^{m_{i_1i_1}^R}\bar{X}_{i_1,b}^R(t)2^{-b} \rfloor \\
              \lfloor \mathrm{sign}(\bar{X}_{i_1}^I(t))h_{i_1i_1}^R\sum_{b=m_{i_1^*i_1}^R+1}^{m_{i_1i_1}^R}\bar{X}_{i_1,b}^I(t)2^{-b} \rfloor \\
            \end{array}
          \right)\label{sum1}\\
          &+ \sum_{k=2}^m
          \left(
            \begin{array}{c}
              \lfloor \mathrm{sign}(\bar{X}_{i_k}^R(t))h_{i_1i_k}^R\sum_{b=1}^{m_{i_1i_k}^R}\bar{X}_{i_k,b}^R(t)2^{-b} \rfloor - \lfloor \mathrm{sign}(\bar{X}_{i_k}^I(t))h_{i_1i_k}^I\sum_{b=1}^{m_{i_1i_k}^I}\bar{X}_{i_k,b}^I(t)2^{-b} \rfloor  \\
              \lfloor \mathrm{sign}(\bar{X}_{i_k}^R(t))h_{i_1i_k}^I\sum_{b=1}^{m_{i_1i_k}^I}\bar{X}_{i_k,b}^R(t)2^{-b} \rfloor + \lfloor \mathrm{sign}(\bar{X}_{i_k}^I(t))h_{i_1i_k}^R\sum_{b=1}^{m_{i_1i_k}^R}\bar{X}_{i_k,b}^I(t)2^{-b} \rfloor\\
            \end{array}
          \right)\label{sum2}\\
          &+\underbrace{\left(
              \begin{array}{c}
                \hat{C}_{i_1}^R(t) \\
                \hat{C}_{i_1}^I(t)\\
              \end{array}
            \right)}_{\hat{C}_{i_1}(t)}
\end{align}
where $i_1^*$ denotes the strongest virtual receiver connected to transmitter $i_1$, i.e., $|h_{i_1^*i_1}|=\max_{j\in\{m+1,...,N\}}\{|h_{i_ji_1}|\}$, and $\hat{C}_{i_1}^R(t)$ and $\hat{C}_{i_1}^I(t)$ can both take values from $\{-1,0,1\}$. Define
\begin{align}
\bar{X}_{i_1^*i_1,S}^R(t)&=\mathrm{sign}(\bar{X}_{i_1}^R(t))\sum_{b=1}^{m_{i_1^*i_1}^R}\bar{X}_{i_1,b}^R(t)2^{-b}\\
\bar{X}_{i_1^*i_1,S}^I(t)&=\mathrm{sign}(\bar{X}_{i_1}^I(t))\sum_{b=1}^{m_{i_1^*i_1}^R}\bar{X}_{i_1,b}^I(t)2^{-b}
\end{align}
Also define the sum of (\ref{sum1}) and (\ref{sum2}) as $\hat{Y}_{i_1,l}(t)$, i.e.,
\begin{align*}
\hat{Y}_{i_1,l}(t)=&\left(
            \begin{array}{c}
              \lfloor \mathrm{sign}(\bar{X}_{i_1}^R(t))h_{i_1i_1}^R\sum_{b=m_{i_1^*i_1}^R+1}^{m_{i_1i_1}^R}\bar{X}_{i_1,b}^R(t)2^{-b} \rfloor \\
              \lfloor \mathrm{sign}(\bar{X}_{i_1}^I(t))h_{i_1i_1}^R\sum_{b=m_{i_1^*i_1}^R+1}^{m_{i_1i_1}^R}\bar{X}_{i_1,b}^I(t)2^{-b} \rfloor \\
            \end{array}
          \right)\\
          &+ \sum_{k=2}^m
          \left(
            \begin{array}{c}
              \lfloor \mathrm{sign}(\bar{X}_{i_k}^R(t))h_{i_1i_k}^R\sum_{b=1}^{m_{i_1i_k}^R}\bar{X}_{i_k,b}^R(t)2^{-b} \rfloor - \lfloor \mathrm{sign}(\bar{X}_{i_k}^I(t))h_{i_1i_k}^I\sum_{b=1}^{m_{i_1i_k}^I}\bar{X}_{i_k,b}^I(t)2^{-b} \rfloor  \\
              \lfloor \mathrm{sign}(\bar{X}_{i_k}^R(t))h_{i_1i_k}^I\sum_{b=1}^{m_{i_1i_k}^I}\bar{X}_{i_k,b}^R(t)2^{-b} \rfloor + \lfloor \mathrm{sign}(\bar{X}_{i_k}^I(t))h_{i_1i_k}^R\sum_{b=1}^{m_{i_1i_k}^R}\bar{X}_{i_k,b}^I(t)2^{-b} \rfloor\\
            \end{array}
          \right)
\end{align*}
Then we have
\begin{align}
\hat{Y}_{i_1}(t)=\hat{Y}_{i_1,u}(t)+\hat{Y}_{i_1,l}(t)+\hat{C}_{i_1}(t)
\end{align}

For receiver $i_1$, starting from Fano's inequality,
\begin{align}
&n(\sum_{j=1}^mR_{i_1i_j}-\epsilon)\\
&\leq I(\mathcal{W}_{i_1}^*;\hat{Y}_{i_1,u}^n,\hat{Y}_{i_1,l}^n,\hat{C}_{i_1}^n,\hat{S}_{i_1}^n|\mathcal{W}_{i_2i_2}^c/\mathcal{W}_{i_1}^*)\\
&=H(\hat{Y}_{i_1,u}^n,\hat{Y}_{i_1,l}^n,\hat{C}_{i_1}^n,\hat{S}_{i_1}^n|\mathcal{W}_{i_2i_2}^c/\mathcal{W}_{i_1}^*)-H(\hat{Y}_{i_1,u}^n,\hat{Y}_{i_1,l}^n,\hat{C}_{i_1}^n,\hat{S}_{i_1}^n|\mathcal{W}_{i_2i_2}^c)\\
&=H(\hat{S}_{i_1}^n|\mathcal{W}_{i_2i_2}^c/\mathcal{W}_{i_1}^*)+H(\hat{Y}_{i_1,u}^n,\hat{Y}_{i_1,l}^n,\hat{C}_{i_1}^n|\hat{S}_{i_1}^n,\mathcal{W}_{i_2i_2}^c/\mathcal{W}_{i_1}^*)
-H(\hat{Y}_{i_1,u}^n,\hat{Y}_{i_1,l}^n,\hat{C}_{i_1}^n|\mathcal{W}_{i_2i_2}^c)\\
&~~~-H(\hat{S}_{i_1}^n|\hat{Y}_{i_1,u}^n,\hat{Y}_{i_1,l}^n,\hat{C}_{i_1}^n,\mathcal{W}_{i_2i_2}^c)\\
&\leq H(\hat{S}_{i_1}^n|W_{i_2i_1},W_{i_3i_1},...,W_{i_mi_1})+H(\hat{Y}_{i_1,u}^n,\hat{Y}_{i_1,l}^n,\hat{C}_{i_1}^n|\hat{S}_{i_1}^n)-H(\hat{Y}_{i_1,u}^n,\hat{Y}_{i_1,l}^n,\hat{C}_{i_1}^n|\mathcal{W}_{i_2i_2}^c)\label{G_R1}
\end{align}
where the last inequality follows from the fact that dropping conditioning does not reduce entropy. Now consider the last term in (\ref{G_R1}),
\begin{align}
&H(\hat{Y}_{i_1,u}^n,\hat{Y}_{i_1,l}^n,\hat{C}_{i_1}^n|\mathcal{W}_{i_2i_2}^c)\\
&=H(\hat{Y}_{i_1,u}^n|\mathcal{W}_{i_2i_2}^c)+H(\hat{Y}_{i_1,l}^n,\hat{C}_{i_1}^n|\hat{Y}_{i_1,u}^n,\mathcal{W}_{i_2i_2}^c) \label{G_T1}\\
&\geq H(\hat{Y}_{i_1,u}^n|\mathcal{W}_{i_2i_2}^c)+H(\hat{S}_{i_2}'^n|\hat{Y}_{i_1,u}^n,\mathcal{W}_{i_2i_2}^c)\\
&=H(\bar{X}_{i_1^*i_1,S}^{R~n},\bar{X}_{i_1^*i_1,S}^{I~n}|\mathcal{W}_{i_2i_2}^c)+H(\hat{S}_{i_2}^n|\hat{Y}_{i_1,u}^n,\mathcal{W}_{i_2i_2}^c)\label{G_T2}\\
&\geq H(\hat{Y}_{i_1^*}^n|\mathcal{\widetilde{W}}/\mathcal{W}_{i_1'})+ H(\hat{S}_{i_2}^n|W_{i_1i_2},W_{i_3i_2},...,W_{i_mi_2}) \label{G_T3}
\end{align}
where (\ref{G_T2}) follows Lemma \ref{complex_map}, i.e., both functions $f: \hat{Y}_{i_1,u} \rightarrow \bar{X}_{i_1^*i_1,S}^R \times \bar{X}_{i_1^*i_1,S}^I$ and $f: \hat{S}_{i_2}' \rightarrow \hat{S}_{i_2}$ are bijective.


Plugging (\ref{G_T3}) into (\ref{G_R1}), we have
\begin{equation}
\label{G_R1_1}
\begin{aligned}
n(\sum_{j=1}^mR_{i_1i_j}-\epsilon)\leq& H(\hat{S}_{i_1}^n|W_{i_2i_1},W_{i_3i_1},...,W_{i_mi_1})+H(\hat{Y}_{i_1,u}^n,\hat{Y}_{i_1,l}^n,\hat{C}_{i_1}^n|\hat{S}_{i_1}^n)\\
&-H(\hat{S}_{i_2}^n|W_{i_1i_2},W_{i_3i_2},...,W_{i_mi_2})-H(\hat{Y}_{i_1^*}^n|\mathcal{\widetilde{W}}/\mathcal{W}_{i_1'})
\end{aligned}
\end{equation}

Then consider the degraded BC comprised of the transmitter $i_1$ and the virtual receivers $\{R'_{i_{m+1}i_1},R'_{i_{m+2}i_1},...,R'_{i_Ni_1}\}$. Since $R'_{i_1^*}$ is the strongest receiver which can decode all the messages from transmitter $i_1$ to all the connected virtual receivers, we have
\begin{align}
n(\sum_{j=m+1}^NR_{i_ji_1}-\epsilon)\leq& I(\mathcal{W}_{i_1'};\hat{Y}_{i_1^*}^n|\mathcal{\widetilde{W}}/\mathcal{W}_{i_1'})\\
=&H(\hat{Y}_{i_1^*}^n|\mathcal{\widetilde{W}}/\mathcal{W}_{i_1'})\label{G_BC}
\end{align}

Adding (\ref{G_R1_1}) and (\ref{G_BC}), we have
\begin{equation}
\begin{aligned}\label{eq1}
&n(\sum_{j=1}^mR_{i_1i_j}+\sum_{j=m+1}^NR_{i_ji_1}-\epsilon)\\
&\leq  H(\hat{S}_{i_1}^n|W_{i_2i_1},W_{i_3i_1},...,W_{i_mi_1})+H(\hat{Y}_{i_1,u}^n,\hat{Y}_{i_1,l}^n,\hat{C}_{i_1}^n|\hat{S}_{i_1}^n)-H(\hat{S}_{i_2}^n|W_{i_1i_2},W_{i_3i_2},...,W_{i_mi_2})
\end{aligned}
\end{equation}

Similarly, for $\forall k\in\{2,3,...,m-1\}$ we can obtain
\begin{equation}
\begin{aligned}\label{eq2}
&n(\sum_{j=1}^mR_{i_ki_j}+\sum_{j=m+1}^NR_{i_ji_k}-\epsilon)\\
&\leq  H(\hat{S}_{i_k}^n|W_{i_1i_k},...,W_{i_{k-1}i_k},W_{i_{k+1}i_k},...,W_{i_mi_k})+H(\hat{Y}_{i_k,u}^n,\hat{Y}_{i_k,l}^n,\hat{C}_{i_k}^n|\hat{S}_{i_k}^n)\\
&~~~-H(\hat{S}_{i_{k+1}}^n|W_{i_1i_{k+1}},...,W_{i_{k}i_{k+1}},W_{i_{k+2}i_{k+1}}...,W_{i_mi_{k+1}})
\end{aligned}
\end{equation}
And from receiver $i_m$ and the degraded BC comprised of transmitter $i_m$ and all its connected virtual receivers, we have
\begin{equation}
\begin{aligned}\label{eq3}
&n(\sum_{j=1}^mR_{i_mi_j}+\sum_{j=m+1}^NR_{i_ji_m}-\epsilon)\\
&\leq  H(\hat{S}_{i_m}^n|W_{i_1i_m},W_{i_2i_m}...,W_{i_{m-1}i_m})+H(\hat{Y}_{i_m,u}^n,\hat{Y}_{i_m,l}^n,\hat{C}_{i_m}^n|\hat{S}_{i_m}^n)\\
&~~~-H(\hat{S}_{i_1}^n|W_{i_2i_{1}},W_{i_3i_1}...,W_{i_mi_{1}})
\end{aligned}
\end{equation}

Adding all the terms in (\ref{eq1}), (\ref{eq2}) and (\ref{eq3}) together and applying the same argument used in Example \ref{comx-ex}, we have
\begin{align}
n(R_{\Sigma,D}-\epsilon)&\leq \sum_{k=1}^m H(\hat{Y}_{i_k,u}^n,\hat{Y}_{i_k,l}^n,\hat{C}_{i_k}^n|\hat{S}_{i_k}^n)\\
&\leq \sum_{k=1}^m\sum_{t=1}^n[H(\hat{Y}_{i_k,u}(t)|\hat{S}_{i_k}(t))+H(\hat{Y}_{i_k,l}(t)|\hat{S}_{i_k}(t))+H(\hat{C}_{i_k}(t))]\\
&\leq n\sum_{k=1}^m2[m_{i_ki_k}-m_{i_{k-1}i_k}]+\mathrm{constant}\label{G_F2}\\
&\leq n\sum_{k=1}^m[\alpha_{i_ki_k}-\alpha_{i_{k-1}i_k}]\log_2 P+\mathrm{constant}
\end{align}
where (\ref{G_F2}) holds since $m_{i_ki_j}-1 \leq \max\{m_{i_ki_j}^R,m_{i_ki_j}^I\}\leq m_{i_ki_j} $, and also note  the modulo-$m$ arithmetic is implicitly used on user indices, e.g., $i_{0}=i_m$.

Finally according to Lemma \ref{lemma_const}, we can obtain the desired GDoF cycle bound for the original complex Gaussian $X$ channel through some simple manipulations,
\begin{align}
\sum_{j=1}^{m}\bar{d}_{i_j}\leq \sum_{j=1}^{m}(\alpha_{i_ji_j}-\alpha_{i_{j-1}i_j})
\end{align}
and hence complete the whole proof.

\section{Conclusion}
In this paper, we extend the optimality of TIN to more general classes of message sets. The main result is that for the TIN-optimal $K$-user interference channel, even if the message set expands to include the $X$ setting where each transmitter has one independent message to each receiver, operating the new channel as the original interference channal and treating interference as noise at each receiver is still optimal to achieve the sum channel capacity to within a constant gap. Furthermore, the optimality of TIN for the general $M\times N$ $X$ channel is also demonstrated.

We conclude with a comment on the necessity of the optimality conditions. In \cite{Geng_TIN_opt} it is conjectured that (\ref{cond_ic}) is also necessary for TIN to be  optimal for the \emph{entire GDoF region} except for a set of channel gain values with measure zero. However,  note that no claim is made for the necessity of  condition (\ref{cond_ic}) for the optimality of TIN for \emph{sum-GDoF}. In fact it is easy to see that (\ref{cond_ic})  is not necessary for the sum-GDoF optimality of TIN. For example,  consider the 2-user interference channel with $\alpha_{11}>\alpha_{12}+\alpha_{21}>\alpha_{22}$, which violates (\ref{cond_ic}), and whose optimal sum-GDoF value, $\alpha_{11}$ (as shown in \cite{Tse_GDoF}),  is trivially achieved by activating only user 1. Similarly, since our focus is only on sum-GDoF, the optimality conditions are only  sufficient, but not necessary.

\appendix
\section{Appendix}

\subsection{Proof of Lemma \ref{lemma_const}} \label{appendix-lemma}

We begin with the general complex Gaussian $M\times N$ $X$ channel, and convert it to the corresponding truncated deterministic channel step-by-step. In each step, we show that only a loss of constant bits is introduced. Here we follow the similar steps used in \cite{Bresler_deterministic}. For the sake of simplicity, we define $\mathcal{W}_{i}^{\star}=\{W_{i1},W_{i2},...,W_{iM}\}$, and we suppress the time index $t$ if no confusion would be caused.


\begin{itemize}
\item Step 1: \textbf{Average power constraint to peak power constraint.} Recall that in the original complex Gaussian channels, by scaling the output, we set
    \begin{align*}
    E[|X_i(t)|^2]\leq 2, ~Z_i(t) \sim \mathcal{CN}(0,2)
    \end{align*}
Then for each input $X_i=X_i^R+jX_i^I$, we truncate both the real and imaginary parts to satisfy the peak power constraint of $1$. Define the part of input $X_i^R$ that exceeds the peak power constraint as
\begin{align*}
\tilde{X}_i^R=\lfloor X_i^R \rfloor = \mathrm{sign}(X_i^R)\sum_{b=-\infty}^0X_{i,b}^R2^{-b}
\end{align*}
and the remaining signal as
\begin{align*}
\bar{X}_i^R=X_i^R-\tilde{X}_i^R=\mathrm{sign}(X_i^R)\sum_{b=1}^{\infty}X_{i,b}^R2^{-b}
\end{align*}
For the imaginary part of the input, we have the similar definitions for $X_i^I$ with I replacing R. Then $\bar{X}_i^R$ and $\bar{X}_i^I$ satisfy the peak power constraint. Letting $\bar{Y}_i$ be the output of receiver $i$ due to the truncated input $\bar{X}_i=\bar{X}_i^R+j\bar{X}_i^I$, and $\tilde{Y_i}$ be the difference between $Y_i$ and $\bar{Y_i}$, for each receiver $i\in\{1,2,...,N\}$ we have
\begin{align}
I(\mathcal{W}_{i}^{\star};Y_i^n)&\leq I(\mathcal{W}_{i}^{\star};\bar{Y}_i^n,\tilde{Y}_i^n)\\
&=I(\mathcal{W}_{i}^{\star};\bar{Y}_i^n)+I(\mathcal{W}_{i}^{\star};\tilde{Y}_i^n|\bar{Y}_i^n)\\
&\leq I(\mathcal{W}_{i}^{\star};\bar{Y}_i^n)+H(\tilde{Y}_i^n)\\
&\leq I(\mathcal{W}_{i}^{\star};\bar{Y}_i^n)+\sum_{j=1}^MH(\tilde{X}_j^n)\\
&\leq I(\mathcal{W}_{i}^{\star};\bar{Y}_i^n)+n\times \mathrm{constant}
\end{align}
where the last inequality comes from Lemma 6 in \cite{Bresler_deterministic}.

\item Step 2: \textbf{Truncate signals at noise level and remove noise.}
Recall $\lfloor \log_2 |h_{ik}^R|\rfloor = m_{ik}^R$ and $\lfloor \log_2 |h_{ik}^I|\rfloor = m_{ik}^I$. We have
\begin{align*}
\hat{Y}_i
=&\sum_{k=1}^M \big[(\lfloor \mathrm{sign}(X_{k}^R)h_{ik}^R\sum_{b=1}^{m_{ik}^R}X_{k,b}^R2^{-b}\rfloor - \lfloor \mathrm{sign}(X_{k}^I)h_{ik}^I\sum_{b=1}^{m_{ik}^I}X_{k,b}^I2^{-b}\rfloor)\\
&+j(\lfloor \mathrm{sign}(X_{k}^R)h_{ik}^I\sum_{b=1}^{m_{ik}^I}X_{k,b}^R2^{-b}\rfloor + \lfloor \mathrm{sign}(X_{k}^I)h_{ik}^R\sum_{b=1}^{m_{ik}^R}X_{k,b}^I2^{-b}\rfloor)\big]
\end{align*}

Define
\begin{align*}
\varepsilon_i =& \bar{Y}_i-\hat{Y_i}\\
=&\sum_{k=1}^M \Big\{\big[\mathrm{sign}(X_{k}^R)h_{ik}^R\sum_{b=m_{ik}^R+1}^{\infty}X_{k,b}^R2^{-b}-\mathrm{sign}(X_{k}^I)h_{ik}^I\sum_{b=m_{ik}^I+1}^{\infty}X_{k,b}^I2^{-b}\\
&+\mathrm{frac}(\mathrm{sign}(X_{k}^R)h_{ik}^R\sum_{b=1}^{m_{ik}^R}X_{k,b}^R2^{-b})-\mathrm{frac}(\mathrm{sign}(X_{k}^I)h_{ik}^I\sum_{b=1}^{m_{ik}^I}X_{k,b}^I2^{-b})\big]\\
&+j\big[\mathrm{sign}(X_{k}^R)h_{ik}^I\sum_{b=m_{ik}^I+1}^{\infty}X_{k,b}^R2^{-b}+\mathrm{sign}(X_{k}^I)h_{ik}^R\sum_{b=m_{ik}^R+1}^{\infty}X_{k,b}^I2^{-b}\\
&+\mathrm{frac}(\mathrm{sign}(X_{k}^R)h_{ik}^I\sum_{b=1}^{m_{ik}^I}X_{k,b}^R2^{-b})+\mathrm{frac}(\mathrm{sign}(X_{k}^I)h_{ik}^R\sum_{b=1}^{m_{ik}^R}X_{k,b}^I2^{-b})\big]\Big\}+Z_i\\
=&\sum_{k=1}^M \hat{X}_k+Z_i
\end{align*}
where $\mathrm{frac}(x)$ denotes the fractional part of $x$. Also note
\begin{align*}
|h_{ik}^R\sum_{b=m_{ik}^R+1}^{\infty}X_{k,b}^R2^{-b}|\leq 2^{m_{ik}^R+1}2^{-(m_{ik}^R)}=2
\end{align*}
Similarly, we have
\begin{align*}
|h_{ik}^I\sum_{b=m_{ik}^I+1}^{\infty}X_{k,b}^I2^{-b}|&\leq 2\\
|h_{ik}^I\sum_{b=m_{ik}^I+1}^{\infty}X_{k,b}^R2^{-b}|&\leq 2\\
|h_{ik}^R\sum_{b=m_{ik}^R+1}^{\infty}X_{k,b}^I2^{-b}|&\leq 2
\end{align*}

Finally, we can obtain
\begin{align}
I(\mathcal{W}_{i}^{\star};\bar{Y}_i^n)&\leq I(\mathcal{W}_{i}^{\star};\hat{Y}_i^n,\varepsilon_i^n)\\
&=I(\mathcal{W}_{i}^{\star};\hat{Y}_i^n)+I(\mathcal{W}_{i}^{\star};\varepsilon_i^n|\hat{Y}_i^n)\\
&=I(\mathcal{W}_{i}^{\star};\hat{Y}_i^n)+h(\varepsilon_i^n|\hat{Y}_i^n)-h(\varepsilon_i^n|\hat{Y}_i^n,\mathcal{W}_{i}^{\star})\\
&\leq I(\mathcal{W}_{i}^{\star};\hat{Y}_i^n) + h(\varepsilon_i^n) - h(Z_i^n)\\
&=I(\mathcal{W}_{i}^{\star};\hat{Y}_i^n) + I(\hat{X}_1^n,\hat{X}_2^n,...,\hat{X}_M^n;\varepsilon_i^n)\\
&\leq I(\mathcal{W}_{i}^{\star};\hat{Y}_i^n) + n\times \mathrm{constant}
\end{align}
in which the last inequality is due to the fact that $\hat{X}_1,\hat{X}_2,...,\hat{X}_M\mapsto \varepsilon_i$ forms a complex Gaussian MAC with a finite SNR independent of $P$ for each transmitter \cite{Bresler_deterministic}.  \hfill $\Box$

\end{itemize}

\subsection{Proof of Lemma \ref{complex_map}} \label{appendix-lemma2}
For notation brevity, we define
\begin{align}
\breve{X}_R\triangleq&\mathrm{sign}(\bar{X}_{i_j}^R(t))\sum_{b=1}^{\max\{m_{i_{j-1}i_j}^R,m_{i_{j-1}i_j}^I\}}\bar{X}_{i_j,b}^R(t)2^{-b} \\
\breve{X}_I\triangleq&\mathrm{sign}(\bar{X}_{i_j}^I(t))\sum_{b=1}^{\max\{m_{i_{j-1}i_j}^R,m_{i_{j-1}i_j}^I\}}\bar{X}_{i_j,b}^I(t)2^{-b} \\
\breve{S}_R\triangleq&\underbrace{\lfloor \mathrm{sign}(\bar{X}_{i_j}^R(t))h_{i_{j-1}i_j}^R\sum_{b=1}^{m_{i_{j-1}i_j}^R}\bar{X}_{i_j,b}^R(t)2^{-b} \rfloor}_{\breve{S}_{R,1}} - \underbrace{\lfloor \mathrm{sign}(\bar{X}_{i_j}^I(t))h_{i_{j-1}i_j}^I\sum_{b=1}^{m_{i_{j-1}i_j}^I}\bar{X}_{i_j,b}^I(t)2^{-b} \rfloor}_{\breve{S}_{R,2}} \\
\breve{S}_I\triangleq&\underbrace{\lfloor \mathrm{sign}(\bar{X}_{i_j}^R(t))h_{i_{j-1}i_j}^I\sum_{b=1}^{m_{i_{j-1}i_j}^I}\bar{X}_{i_j,b}^R(t)2^{-b} \rfloor}_{\breve{S}_{I,1}} + \underbrace{ \lfloor \mathrm{sign}(\bar{X}_{i_j}^I(t))h_{i_{j-1}i_j}^R\sum_{b=1}^{m_{i_{j-1}i_j}^R}\bar{X}_{i_j,b}^I(t)2^{-b} \rfloor}_{\breve{S}_{I,2}}
\end{align}
Note $(\breve{X}_R,\breve{X}_I)$ and $(\breve{S}_R,\breve{S}_I)$ can be seen as the input and output of the deterministic channel, respectively. Obviously, given one input $(\breve{X}_R,\breve{X}_I)$, we can only produce one output $(\breve{S}_R,\breve{S}_I)$. Next, we prove the other direction by contradiction. We assume there exist two different inputs ($\breve{X}_R^*,\breve{X}_I^*$) and ($\breve{X}_R^{**},\breve{X}_I^{**}$) that can generate the same output, i.e., $(\breve{S}_R^*,\breve{S}_I^*) = (\breve{S}_R^{**},\breve{S}_I^{**})$. In the following, without loss of generality, we assume
\begin{align}
|h_{i_{j-1}i_j}^R|\geq |h_{i_{j-1}i_j}^I| \Rightarrow m_{i_{j-1}i_j}^R\geq m_{i_{j-1}i_j}^I.
\end{align}

We first consider the case where $\mathrm{sign}(h_{i_{j-1}i_j}^R) = \mathrm{sign}(h_{i_{j-1}i_j}^I)$. For the term $\breve{S}_R$, we have the following subcases:
\begin{itemize}
\item  $\breve{S}_{R,1}^*=\breve{S}_{R,1}^{**}$ and $\breve{S}_{R,2}^*=\breve{S}_{R,2}^{**}$. In this case, for the term $\breve{S}_I$, if $|h_{i_{j-1}i_j}^R|> |h_{i_{j-1}i_j}^I|$, since ($\breve{X}_R^*,\breve{X}_I^*$) and ($\breve{X}_R^{**},\breve{X}_I^{**}$) are different,  we have $\breve{S}_{I,1}^*=\breve{S}_{I,1}^{**}$ and $\breve{S}_{I,2}^*\neq\breve{S}_{I,2}^{**}$, which contradicts the assumption that $(\breve{X}_R^*,\breve{X}_I^*$) and ($\breve{X}_R^{**},\breve{X}_I^{**}$) generate the same output;  if $|h_{i_{j-1}i_j}^R| = |h_{i_{j-1}i_j}^I|$, since $(\breve{X}_R^*,\breve{X}_I^*$) and ($\breve{X}_R^{**},\breve{X}_I^{**}$) generate the same ($\breve{S}_R,\breve{S}_I$), we have $(\breve{X}_R^*,\breve{X}_I^*) = (\breve{X}_R^{**},\breve{X}_I^{**})$, which contradicts the assumption that ($\breve{X}_R^*,\breve{X}_I^*$) and ($\breve{X}_R^{**},\breve{X}_I^{**}$) are different.
\item  $\breve{S}_{R,1}^*>\breve{S}_{R,1}^{**}$ and $\breve{S}_{R,2}^*>\breve{S}_{R,2}^{**}$. In this case, for the term $\breve{S}_I$, we have $\breve{S}_{I,1}^*\geq\breve{S}_{I,1}^{**}$ and $\breve{S}_{I,2}^*>\breve{S}_{I,2}^{**}$, which contradicts the assumption that $(\breve{X}_R^*,\breve{X}_I^*$) and ($\breve{X}_R^{**},\breve{X}_I^{**}$) generate the same output.
\item  $\breve{S}_{R,1}^*<\breve{S}_{R,1}^{**}$ and $\breve{S}_{R,2}^*<\breve{S}_{R,2}^{**}$. In this case, for the term $\breve{S}_I$, we have $\breve{S}_{I,1}^*\leq\breve{S}_{I,1}^{**}$ and $\breve{S}_{I,2}^*<\breve{S}_{I,2}^{**}$, which contradicts the assumption that $(\breve{X}_R^*,\breve{X}_I^*$) and ($\breve{X}_R^{**},\breve{X}_I^{**}$) generate the same output.
\end{itemize}

For the other case where $\mathrm{sign}(h_{i_{j-1}i_j}^R) = -\mathrm{sign}(h_{i_{j-1}i_j}^I)$, we can follow the same argument to get the same conclusion. \hfill $\Box$

\end{document}